\documentclass[12pt]{iopart}
\usepackage[utf8]{inputenc}
\usepackage{graphicx}
\usepackage{iopams}
\usepackage{dsfont}

\def\be{\begin{equation}}
\def\ee{\end{equation}}
\def\ba{\begin{eqnarray}}
\def\ea{\end{eqnarray}}

\def\ket#1{|{#1}\rangle}
\def\bra#1{\langle{#1}|}
\def\braket#1#2{\langle{#1}|{#2}\rangle}

\def\kket#1{|{#1}\rangle\!\rangle}

\def\bigmoyenne#1{\big\langle{#1}\big\rangle}

\def\exp#1{\mathop{\rm exp}\big(#1\big)}
\def\trace{\mathop{\rm Tr}\nolimits}
\def\identity{\mathds{1}}

\begin{document}
\title[Introduction to MPS algorithms for Lindblad equation]
{A first introduction to Matrix Product State algorithms
for the integration of Lindblad equation}
\author{Christophe Chatelain}
\address{Universit\'e de Lorraine, CNRS, LPCT, F-54000 Nancy, France}
\ead{christophe.chatelain@univ-lorraine.fr}
\date{\today}

\begin{abstract}
In this introductory review, we present and compare four algorithms
for the numerical integration of the Lindblad equation for one-dimensional
quantum lattice systems. All four methods are based on Matrix Product State
representations and can be viewed as extensions of the Time-Evolving Block
Decimation (TEBD) algorithm to open quantum systems. Two approaches directly
integrate the vectorized Lindblad equation, one of them explicitly enforcing
the positivity of the density matrix. The other two rely on stochastic
unravelings of the Lindblad equation, namely the quantum trajectory and
quantum state diffusion approaches. We discuss the principles, numerical
implementation, accuracy, and computational efficiency of the different
methods, and benchmark them against an exactly solvable free fermion model.
\end{abstract}
\maketitle

\section{Introduction}
The study of open quantum systems is a very active field of research
stimulated by the existence of well-controlled experiments, either with
ultra-cold atoms~\cite{Weiner}, atoms in optical cavity~\cite{Ritsch},
or superconducting circuits~\cite{Krantz}, and by the development of
devices processing quantum information~\cite{Amico}.
From the theoretical point of view, open quantum systems pose several
challenges, one of them being to find efficient numerical algorithms to
simulate their dynamics~\cite{Fazio}. Indeed, a full description of the
dynamics of the system and its environment via the von Neumann equation
	\be {d\rho\over dt}=-i[H,\rho]		\label{vonNeumann}\ee
where $\rho$ is the density matrix and $H$ the Hamiltonian of the whole
system (system and environment), is generally out-of-reach. Instead, one
looks for an effective dynamical equation for the reduced density matrix
	\be \rho_S=\trace_{\rm Env.}\rho\ee
where the trace is taken over the degrees of freedom of the environment.
The establishment of such an equation requires to make some assumptions.
Assuming a weak interaction (Born approximation), a Markovian dynamics
and performing the rotating-wave approximation, one obtains the celebrated
Lindblad (or Gorini-Kossakowski-Sudarshan-Lindblad) equation~\cite{Breuer,
Manzano,Stefanini}
	\be {d\rho_S\over dt}=-i[H_S,\rho_S]
	+\sum_k \gamma_k \Big[L_k\rho_S L_k^+
	-{1\over 2}\{L_k^+L_k,\rho_S\}\Big].		\label{Lindblad}\ee
While the first term of the r.h.s. leads to a unitary dynamics, the terms
involving the jump operators $L_k$ cause dissipation in the system and a
loss of quantum coherence. As will be seen in the following, it is often
convenient to rewrite Lindblad equation as
	\be {d\rho_S\over dt}=-iH_{\rm eff}\rho_S+i\rho_SH_{\rm eff}^+
	+\sum_k \gamma_k L_k\rho_S L_k^+		\label{LindbladEff}\ee
with the non-hermitian effective Hamiltonian
	\be H_{\rm eff}=H-{i\over 2}\sum_k \gamma_k L_k^+L_k.
	\label{EffHamiltonian}\ee

This introductory article focuses on algorithms for the numerical integration
of the Lindblad equation (\ref{Lindblad}) for one-dimensional quantum systems.
It is intended to students or young researchers who want to learn and
experiment by themselves by coding their own simulation. Readers whose goal
is to become rapidly productive are advised to have a look to ready-to-use
libraries, as for instance {\tt TeNPy}~\cite{Hauschild} or
{\tt TensorMixedStates}~\cite{Houdayer}. More advanced and exhaustive
reviews, including in particular a discussion on two-dimensional systems,
can be found in the literature~\cite{Jaschke,Weimer}.
The algorithms will be illustrated
by simulations of a simple model of free fermions. The main interest of the
latter is that Lindblad equation admits an exact solution, which will be
presented in the second section and will be used in the rest of the
paper to determine the systematic deviation of the numerical results.
Before discussing algorithms for Lindblad equation, the Matrix Product
States (MPS) representation of the wavefunction and the Time-Evolving
Block Decimation (TEBD) algorithm to simulate a unitary evolution will
be briefly reviewed in the rest of the second section.
The third section will be devoted to algorithms based on a vectorization
of Lindblad equation by means of Choi-Jamielowski isomorphism. While the
formulation in terms of Matrix Product Density Operator (MPDO) does not
preserve the positivity of the density matrix,
a simple extension based on a local purification of the density matrix
will be presented. In the fourth section, algorithms relying on a stochastic
interpretation of the density matrix and its estimate by a Monte Carlo
simulation will be discussed. The action of the jump operators will be either
implemented as quantum jumps or as a continuous diffusion term. Conclusions
and perspectives will follow.

\section{The model and MPS algorithms for the unitary evolution}
In this section, the model that will be used to test the various algorithms
presented in this paper is first described. The exact solution of Lindblad
equation for correlation functions is derived. It will serve as a reference
for analyzing the systematic deviations of the numerical algorithms.
In the rest of the section, the decomposition of the wavefunction as
a Matrix Product State (MPS) and the TEBD algorithm for the integration
of the von Neumann equation (\ref{vonNeumann}) are briefly reviewed
in order to provide a coherent presentation.

\subsection{The model}\label{SecFermion}
Throughout this paper, a simple one-dimensional tight-binding model of free
fermions hopping on a chain with a source on the left and a sink on the right
will be considered. The Hilbert space of a chain of $N$ sites is the tensor
product ${\cal H}^{\otimes N}$ where the single-site Hilbert space is
${\cal H}={\mathbb C}^d$ with $d=2$. The Hamiltonian of the system is chosen
as
	\be H_S=-\sum_{i=1}^{N-1} \big(c_{i+1}^+c_i+c_i^+c_{i+1}\big)
	\label{Hamiltonian}\ee
where $c_i^+$ (resp. $c_i$) is the creation (resp. annihilation)
fermionic operator acting on site~$i$
	\be c_i^+=\underbrace{\identity\otimes\identity\otimes\ldots\otimes
	\identity}_{(i-1)\ {\rm times}}\otimes\ \! c^+\otimes\underbrace{
	\identity\otimes\ldots\otimes\identity}_{(N-i)\ {\rm times}}.\ee
These operators satisfy the anti-commuting algebra $\{c_i,c_j^+\}=\delta_{i,j}$.
The environment is assumed to be a reservoir that injects particles on the left
site of the chain and removes particles on the right one. This amounts to introduce
in Lindblad equation the two jump operators $L_1=c_1^+$ and $L_2=c_N$. For
simplicity, the same coupling $\gamma=\gamma_1=\gamma_2$ will be considered.
Other configurations, with two sources ($L_1=c_1^+$ and $L_2=c_N^+$) or
with a source and dephasing ($L_i=c_i^+c_i$), have been considered. Our choice
was motivated by the fact that it was the hardest to simulate numerically.
The system is initially prepared in the vacuum state $\ket 0$, i.e. $\rho_S(0)
=\ket 0\bra 0$. Note that this system is equivalent, under a Jordan-Wigner
transformation, to a XX spin-$1/2$ chain of Hamiltonian~\cite{Santoro}
	\be H_S=-{1\over 2}\sum_{i=1}^{N-1} \big(\sigma^x_i\sigma^x_{i+1}
	+\sigma^y_i\sigma^y_{i+1}\big)\ee
but with the Lindblad jump operators $L_1=\sigma_1^+$ and $L_2=\prod_{i=1}^{N-1}
(-\sigma^z_i)\ \!\sigma_N^-$~\cite{Yamanaka}.
\\

The main interest of this free-fermion model is that Lindblad
equation~(\ref{Lindblad}) leads to a closed set of linear equations
for the correlations functions $C_{i,j}(t)=\bigmoyenne{c_i^+(t)c_j(t)}$.
The equation of motion of $C_{i,j}$ follows from Lindblad equation (\ref{Lindblad}):
	\ba\fl {dC_{i,j}\over dt}=\trace c_i^+c_j{d\rho_S\over dt}
	=\trace c_i^+c_j\Big[-i[H_S,\rho_S]
	&+&\gamma\Big(c_1^+\rho_S c_1-{1\over 2}c_1c_1^+\rho_S
	-{1\over 2}\rho_S c_1c_1^+\Big)\Big.			\nonumber\\
	\Big. &+&\gamma\Big(c_N\rho_S c_N^+-{1\over 2}c_N^+c_N\rho_S
	-{1\over 2}\rho_S c_N^+c_N\Big)\Big].			\ea
The property of cyclicity of the trace allows all terms to be recast as
averages, for instance $\trace c_i^+c_j[H_S,\rho_S]=\bigmoyenne{[c_i^+c_j,
H_S]}$. Using the anti-commutation relations of the fermionic operators,
one obtains
	\ba {dC_{i,j}\over dt}&=&i\big(C_{i,j+1}+C_{i,j-1}-C_{i+1,j}-C_{i-1,j}\big)
	+\gamma\delta_{i,1}\delta_{j,1}		\nonumber\\
	&&\quad-{\gamma\over 2}\big(C_{i,1}\delta_{j,1}+C_{1,j}\delta_{i,1}\big)
	-{\gamma\over 2}\big(C_{i,N}\delta_{j,N}+C_{N,j}\delta_{i,N}\big)	\ea
where the notations $C_{i,0}=C_{i,N+1}=C_{0,j}=C_{N+1,j}=0$ have been
introduced to allow for an equation valid for any $(i,j)$. In matrix
form, the equation of motion reads
	\be{dC\over dt}=-iH^{\rm eff}C+iC[H^{\rm eff}]^++\Gamma	\label{eqCorr}\ee
with $H_{i,j}^{\rm eff}=\delta_{j,i+1}+\delta_{j,i-1}-{i\gamma\over 2}
\big(\delta_{i,1}\delta_{j,1}+\delta_{N,1}\delta_{j,N}\big)$
and $\Gamma_{i,j}=\gamma\delta_{i,1}\delta_{j,1}$. Note the appearance of
a non-hermitian Hamiltonian $H^{\rm eff}$ as in equation (\ref{LindbladEff}).
The inhomogeneous first-order differential equation (\ref{eqCorr}) admits
the solution
	\be C(t)=e^{-iH^{\rm eff}t}C(0)e^{i[H^{\rm eff}]^+t}
	+\int_0^t e^{-iH^{\rm eff}(t-s)}\Gamma e^{i[H^{\rm eff}]^+(t-s)}ds.
	\label{SoleqCorr}\ee
The system being initially prepared in the vacuum state, the correlations
$C(0)$ vanish. The density profile $\bigmoyenne{n_i(t)}$ is given by the
diagonal elements $C_{ii}(t)$. From the continuity equation
	\be{d\over dt}\bigmoyenne{n_i(t)}=-i\bigmoyenne{[c_i^+c_i,H]}
	=\bigmoyenne{j_{i}}-\bigmoyenne{j_{i+1}}\ee
one can extract the expression of the average current in the chain
	\be \bigmoyenne{j_i}=i\big(\bigmoyenne{c_i^+c_{i-1}}
	-\bigmoyenne{c_{i-1}^+c_i}\big)=i\big(C_{i,i-1}-C_{i-1,i}\big).\ee
In the rest of the paper, a chain of length $N=24$ is considered.
Average density and current profiles will be compared with the numerical
estimates given by the various algorithms that will be discussed.
Instead of considering the exact solution (\ref{SoleqCorr}), the equation
of motion (\ref{eqCorr}) will be integrated numerically using a time
discretization $\Delta t$ ten times smaller than in the MPS algorithms.

\subsection{Matrix Product States}
A major breakthrough in the context of simulation of quantum one-dimensional
lattice systems was the discovery that states with a sufficiently low
entanglement can be efficiently approximated by Matrix Product States (MPS).
Many reviews have already been published on this topic, see in particular
\cite{Misguich,Orus,Hauschild,Cirac,Banuls}. In this paragraph, only the
strictly necessary material for the rest of the paper is presented.
\\

For the above-discussed free fermion model, the quantum state can be
decomposed as
	\be\ket\psi=\sum_{n_1,n_2,\ldots, n_N=0}^1\psi_{n_1,n_2,\ldots, n_N}
	\ket{n_1}\otimes\ket{n_2}\otimes\ldots\otimes\ket{n_N}
	\label{OccupationBasis}\ee
on the occupation-number basis. Two mathematical tools are particularly
useful in the context of tensor-network algorithms. The first one is the
ability to reshape a tensor of rank $r$ into a tensor of lower rank by
grouping indices. The wavefunction $\psi_{n_1,n_2,\ldots, n_N}$
can be seen as a tensor of rank $N$. Consider a bipartition of the chain
into a left and a right part, for example the first site and the rest
of the chain. By introducing the two indices $n_L=n_1\in\{0,d-1\}$ and
$n_R=\sum_{j=2}^N d^{N-j}n_j\in\{0,1,\ldots, d^{N-1}-1\}$, the wavefunction
can be written as $\psi_{n_L,n_R}$ which can be interpreted as the elements
of a $d\times d^{N-1}$ rectangular matrix. The second important mathematical
tool is the Singular Value Decomposition. The matrix $\Psi$, whose elements
are $\psi_{n_L,n_R}$ introduced above, is not square and, as a consequence,
cannot be diagonalized. However, the following Singular Value Decomposition
	\be\Psi=U\Lambda V^+\label{SVD}\ee
can be performed, where $U$ and $V$ are unitary matrices and $\Lambda$ is a
diagonal matrix whose elements are termed as the singular values. In terms
of the original indices, equation (\ref{SVD}) reads
	\be \psi_{n_1,n_2,\ldots, n_N}=\sum_{a} U_{n_1,a}\Lambda_a
	(V^+)_{a,n_2,\ldots n_N}.\label{SVD2}\ee
While $n_1$ to $n_N$ are physical indices corresponding to fermion densities
at each site of the chain, the summation index $a$ is associated to a bond
between two adjacent sites. It is usually referred to as a virtual index.
Its dimension is the minimum of the dimensions of $n_L$ and $n_R$, in our
case $d$. Note that the singular values $\Lambda_a$ are the coefficients
of the Schmidt decomposition $\ket\psi=\sum_a\Lambda_a\ket{\psi_a^L}
\otimes\ket{\psi_a^R}$ for this bipartition of the system. It follows
that the normalization of the wavefunction requires
	\be\sum_a \Lambda_a^2=1			\label{Normalization}\ee
since the unitarity of $U$ (resp. $V$) implies that $\ket{\psi_a^L}$
(resp. $\ket{\psi_a^R}$) are normalized and orthogonal for
different indices $a$. Moreover, the entanglement entropy associated
to this bipartition is easily estimated from these singular values
as~\cite{Latorre}
	\be S_e=-\sum_a \Lambda_a^2\ln\Lambda_a^2.\ee

\begin{figure}
    \centering
    \includegraphics[width=0.75\textwidth]{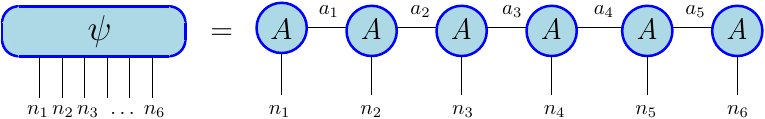}
    \caption{Diagrammatic representation, originally introduced by Penrose,
    of equation (\ref{MPS}) in the case of chain of 6 sites~\cite{Taylor,Evenbly}.
    Tensors are represented as blue boxes. Each one of their legs corresponds
    to an index. Open legs are associated to physical indices $n_i$ while
    internal lines connecting the legs of two tensors are associated
    to virtual indices. The summation over the latter is implicit.
    }\label{fig1}
\end{figure}

The SVD (\ref{SVD2}) has broken the wavefunction into a product.
The process can be iterated: the next step is to reshape $\Lambda_a
(V^+)_{a,n_2,\ldots n_N}$ into a matrix $\Psi^{(2)}_{n_L,n_R}$ by grouping
the indices $a$ and $n_2$ into $n_L$ and $n_3$ to $n_N$ into $n_R$, and
then perform a SVD. After $N-1$ iterations, the wavefunction will take the
form of a Matrix Product State (MPS):
	\be \psi_{n_1,n_2,\ldots, n_N}=\sum_{a_1,\ldots,a_{N-1}}
	A^{n_1}_{\ a_1}A^{n_2}_{\ a_1,a_2}A^{n_3}_{\ a_2,a_3}
	\ldots A^{n_N}_{\ a_{N-1}}.		\label{MPS}\ee
The notation $A^{n_i}$ can be confusing: $A^{n_i}$ is not the $n_i$-th
power of a matrix $A$ but the $n_i$-th element of a set of two matrices.
In the product (\ref{MPS}), either the matrix $A^0$ or $A^1$ is placed at
the $i$-th position depending on the occupation number $n_i$. Except in
the case of a translation-invariant system, these two matrices $A^0$ and
$A^1$ are site-dependent but the more explicit notation $A_i^{n_i}$
is rarely used. Note that $A^{n_1}$ (resp. $A^{n_N}$) is not a matrix
but a row (resp. column) vector. A diagrammatic representation of
equation (\ref{MPS}) is shown on Fig.~\ref{fig1}.
\\

The decomposition (\ref{MPS}) is exact but becomes rapidly intractable
on a computer for long chains. Indeed, the size of the matrices $A^{n_i}$
grow exponentially as $i$ goes to the middle of the chain. The usual
approximation consists in truncating the matrices. This can be obtained
by keeping only the $\chi$ largest singular values during each SVD:
	\be \Psi_{n_L,n_R}\simeq \Psi_{n_L,n_R}^{(\chi)}
	=\sum_{a=1}^\chi U_{n_L,a}\Lambda_a (V^+)_{a,n_R}.\ee
While the wavefunction $\psi_{n_1,n_2,\ldots, n_N}$ comprises $d^N$ complex
numbers, the MPS (\ref{MPS}) requires to store only $Nd\chi^2$ matrix elements.
A convenient estimate of the deviation of $\Psi_{n_L,n_R}^{(\chi)}$ from the
original wavefunction is provided by the square root of the Hilbert-Schmidt
norm. The latter is indeed simply equal to the sum of the squares of the
omitted singular values:
	\ba ||\Psi-\Psi^{(\chi)}||_2
	&=&\sum_{n_L,n_R}\Big|\Psi_{n_L,n_R}-\Psi_{n_L,n_R}^{(\chi)}\Big|^2
	\nonumber\\
	&=&\sum_{n_L,n_R}\bigg|\sum_{a>\chi} U_{n_L,a}\Lambda_a
	(V^+)_{a,n_R}\bigg|^2=\sum_{a>\chi} \Lambda_a^2
	\label{HilbertSchmidtNorm}\ea
where the unitarity of $U$ and $V$ was used. The truncation of the matrices
$A^{n_i}$ yields a good approximation of low-entanglement quantum states,
notably the ground state of one-dimensional gapped Hamiltonians. In the latter,
the singular values are observed to decay exponentially fast which implies
an exponential decay of the truncation error (\ref{HilbertSchmidtNorm}) as
the bond dimension $\chi$ is increased. This property is exploited by the DMRG
algorithm that has been reinterpreted, several years after its introduction,
as a variational method whose variational parameters are the elements
of the $A^{n_i}$ matrices~\cite{DMRG,Schollwoeck,Catarina}.

\begin{figure}
    \centering
    \includegraphics[width=0.7\textwidth]{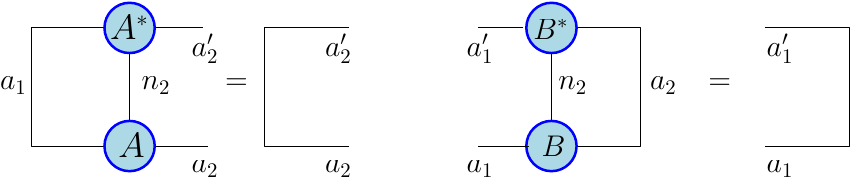}
    \caption{Diagrammatic representation of the gauge conditions
    (\ref{GaugeL}) and (\ref{GaugeR}).}
    \label{fig2}
\end{figure}

\begin{figure}
    \centering
    \includegraphics[width=0.7\textwidth]{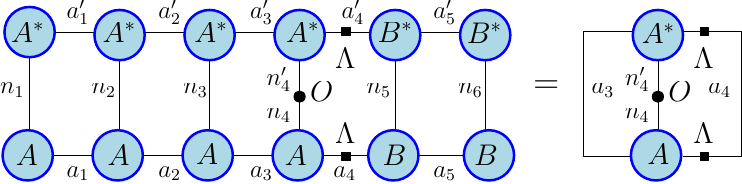}
    \caption{Diagrammatic representation of the expression of the
    quantum average $\bigmoyenne{O}=\bra\psi O\ket\psi$ of an observable
    acting only on site 4 when $\ket\psi$ is decomposed as a mixed MPS
    (\ref{MPSMixte}). The matrix elements $O_{n_4,n_4'}$ of the observable
    are depicted as black dot. Exploiting the left-canonical (\ref{GaugeL}) and
    right-canonical (\ref{GaugeR}) gauge conditions, the tensor network can
    be efficiently contracted leaving a smaller one involving only two
    matrices.}
    \label{fig3}
\end{figure}

The decomposition (\ref{MPS}) is not unique. The transformation
$A^{n_i}\rightarrow A^{n_i}X$ and $A^{n_{i+1}}\rightarrow X^{-1}A^{n_{i+1}}$
leaves indeed the MPS invariant. This property is termed as gauge invariance.
It is often useful to fix the gauge, in particular to compute local quantum
averages. The so-called left-canonical gauge is obtained by imposing the
gauge condition
	\be\sum_{n_i} \big(A^{n_i}\big)^+A^{n_i}=\identity.	\label{GaugeL}\ee
The construction of the MPS with SVD's successively applied from the left of
the chain yields matrices $A^{n_i}$ satisfying this condition, even after
their truncation. In a symmetric way, the right-canonical gauge is obtained
with the condition
	\be\sum_{n_i} B^{n_i}\big(B^{n_i}\big)^+=\identity.	\label{GaugeR}\ee
The matrices are usually denoted $B^{n_i}$ in this case to distinguish them
from those of the left-canonical gauge. These matrices are obtained when
the SVD's are applied successively from the right of the chain.
A diagrammatic representation of the gauge conditions (\ref{GaugeL}) and
(\ref{GaugeR}) is shown on Fig.~\ref{fig2}.
The same quantum state can be written as a product of left-canonical $A^{n_i}$
matrices or as a product of right-canonical $B^{n_i}$ matrices. A mixed
decomposition
	\be \psi_{n_1,n_2,\ldots, n_N}=\sum_{a_1,\ldots,a_{N-1}}
	A^{n_1}_{\ a_1}\ldots A^{n_i}_{\ a_{i-1},a_i}\Lambda_{a_i}
	B^{n_{i+1}}_{\ a_i,a_{i+1}}\ldots B^{n_N}_{\ a_{N-1}}\label{MPSMixte}\ee
is also possible by performing SVD's from both the left and the right of the
chain. Note that it involves an additional term $\Lambda_{a_i}$ that corresponds
to one of the singular values introduced by the successive SVD's of the
quantum state. The location on the chain of this term is called the
orthogonality center. As shown on Fig.~\ref{fig3}, the mixed decomposition
simplifies the numerical computation of quantum averages of local operators.
Note also that the normalization of the wavefunction imposes that
$\sum_a \Lambda_{a_i}=1$. The orthogonality center can be shifted to the
right by performing a SVD decomposition of the matrix
	\be C_{n_L,n_R}=\sum_{a_{i+1}}\Lambda_{a_i}B^{n_{i+1}}_{\ a_i,a_{i+1}}
	B^{n_{i+2}}_{\ a_{i+1},a_{i+2}}			\label{ShiftCenter}\ee
where $n_L=a_id+n_{i+1}$ and $n_R=a_{i+2}d+n_{i+2}$. It is also possible
to use a QR decomposition, which is faster than a SVD.

\subsection{Unitary Time-Evolution algorithms}\label{SecTEBD}
For an isolated system, the time evolution of a pure state is generated
by the Schr\"odinger equation
	\be i{d\over dt}\ket\psi=H_S\ket\psi\ \Leftrightarrow\
	\ket{\psi(t)}=e^{-iH_St}\ket{\psi(0)}.		\label{Schrodinger}\ee
The {\sl Time-Evolving Block Decimation} (TEBD) algorithm allows for the
iterative construction of the time evolution of a MPS approaching
$\ket{\psi(t)}$~\cite{Vidal,Hauschild,Paeckel}. As most algorithms,
it relies on a Suzuki-Trotter decomposition of the evolution operator
	\be e^{-iH_St}=\prod_{i=1}^{n_{\rm iter}} e^{-iH_S\Delta t}\ee
where the time step is $\Delta t=t/n_{\rm iter}$. The evolution
operator $U(\Delta t)$ cannot be determined without approximation.
One can do better than the crude first-order approximation
	\be e^{-iH_S\Delta t}\simeq \identity-iH_S\Delta t.\ee
Our Hamiltonian (\ref{Hamiltonian}) is indeed a sum
	\be H_S=\sum_{i=1}^{N-1} h_{i,i+1}\ee
where $h_{i,i+1}=-\big(c_{i+1}^+c_i+c_i^+c_{i+1}\big)$ acts only on
the neighboring sites $i$ and $i+1$. The local Hamiltonian $h_{i,i+1}$
commute with all other local Hamiltonians, except $h_{i-1,i}$ and $h_{i+1,i+2}$.
As a consequence, the evolution operator $\exp{-iH_S\Delta t}$ cannot
be written as a product $\prod_i \exp{-ih_{i,i+1}\Delta t}$.
However, keeping only the terms with even indices $i$ in the Hamiltonian,
i.e. for $H_e=\sum_{i\ {\rm even}} h_{i,i+1}$, it holds that
	\be e^{-iH_e\Delta t}=\prod_{i\ {\rm even}} e^{-ih_{i,i+1}\Delta t}.\ee
The same relation holds for $H_o=\sum_{i\ {\rm odd}} h_{i,i+1}$ involving
only odd indices $i$. Note that, since $h_{i,i+1}$ acts only on two sites,
it is represented by a $4\times 4$ matrix in the subspace generated by
$\{\ket{n_i}\otimes \ket{n_{i+1}}\}$ so its exponential can be easily
computed numerically either by diagonalization of $h_{i,i+1}$ or,
to good accuracy, by series expansion. The only approximation is finally
	\be\fl e^{-iH_S\Delta t}=e^{-i(H_e+H_o)\Delta t}
	\simeq e^{-iH_e\Delta t}e^{-iH_o\Delta t}
	=\prod_{i\ {\rm even}} e^{-ih_{i,i+1}\Delta t}
	\times \prod_{i\ {\rm odd}} e^{-ih_{i,i+1}\Delta t}\ee
exact only at first order in $\Delta t$. Higher-order schemes in
$\Delta t$, as~\cite{Misguich}
    \be e^{-iH_S\Delta t}=e^{-i(H_e+H_o)\Delta t}
	\simeq e^{-iH_e\Delta t/2}e^{-iH_o\Delta t}e^{-iH_e\Delta t/2}\ee
exact at order $\Delta t^2$, can be considered for a better accuracy.
\\

The TEBD algorithm consists in applying the evolution operators
$\exp{-ih_{i,i+1}\Delta t}$ first on even sites and then on odd sites
at each time step $\Delta t$. Each application of the evolution operators
breaks the MPS structure of the quantum state by coupling two matrices
$A^{n_i}$ and $A^{n_{i+1}}$. The MPS structure should
therefore be restored at each application by means of a SVD. Explicitly,
the application of $\exp{-ih_{i,i+1}\Delta t}$ onto the MPS (\ref{MPS})
amounts to replace $A^{n_i}A^{n_{a+1}}$ by the tensor, often denoted $\Theta$
in the literature,
	\be\Theta_{a_{i-1},n_{i}',a_{i+1},n_{i+1}'}=\sum_{a_{i},n_{i},n_{i+1}}
	\Big[e^{-ih_{i,i+1}\Delta t}\Big]_{n_{i}',n_{i+1}';n_{i},n_{i+1}}
	A^{n_{i}}_{a_{i-1},a_{i}}A^{n_{i+1}}_{a_{i},a_{i+1}}	\label{Theta}\ee
To restore the MPS structure, a SVD of the matrix $\Theta_{n_L,n_R}$
with $n_L=a_{i-1}d+n_{i}'$ and $n_R=a_{i+1}d+n_{i+1}'$ is performed and then
$A^{n_i}$ is replaced by $U\Lambda^{1/2}$ and $A^{n_{i+1}}$ by $\Lambda^{1/2}V^+$.
Again, it will be necessary to truncate the matrices to avoid the
exponential growth of their size. Following (\ref{HilbertSchmidtNorm}),
a truncation error can be defined as
	\be\epsilon=1-\sum_{a>\chi}\Lambda_a^2.		\label{TruncationErr}\ee
Note finally that truncations and time discretization lead to a slow shift
of the norm of the wavefunction (\ref{Normalization}). The simplest way to
overcome this problem is to normalize the MPS after each SVD by imposing
	\be \sum_{a\le\chi}\Lambda_a^1=1.\ee

\begin{figure}
    \centering
    \includegraphics[width=0.52\textwidth]{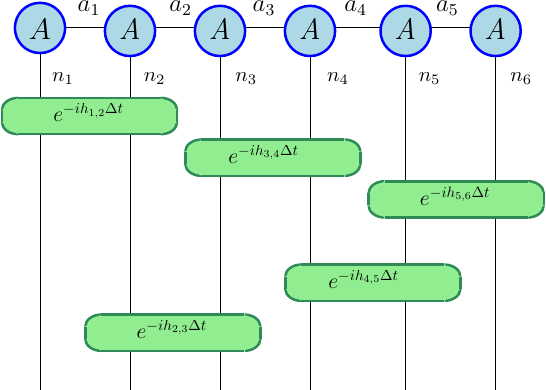}
    \caption{Diagrammatic representation of one iteration of the TEBD
    algorithm corresponding to the unitary time evolution of the MPS
    during a time step $\Delta t$.}
    \label{fig4}
\end{figure}

While the above procedure preserves the MPS structure of the quantum state,
the gauge condition (\ref{GaugeL}) is violated, which will make more
cumbersome the estimate of quantum averages.
To preserve the gauge, one should consider instead the mixed decomposition
(\ref{MPSMixte}). The evolution operator should be applied on the two sites
at the right (or left) of the orthogonality center. The latter being on site
$i$ in the expression (\ref{MPSMixte}), the evolution operator
$\exp{-ih_{i+1,i+2}\Delta t}$ is applied, which amounts to replace
$\Lambda B^{n_{i+1}} B^{n_{i+2}}$ by the tensor
	\be\fl \Theta_{a_i,n_{i+1}',a_{i+2},n_{i+2}'}=\sum_{a_{i+1},n_{i+1},n_{i+2}}
	\Big[e^{-ih_{i+1,i+2}\Delta t}\Big]_{n_{i+1}',n_{i+2}';n_{i+1},n_{i+2}}
	\Lambda_{a_i}B^{n_{i+1}}_{a_i,a_{i+1}}B^{n_{i+2}}_{a_{i+1},a_{i+2}}\ee
To restore the MPS structure, a SVD of the matrix $\Theta_{n_L,n_R}$
with $n_L=a_id+n_{i+1}'$ and $n_R=a_{i+2}d+n_{i+2}'$ is performed. $\Theta$
is therefore replaced by $U\Lambda'V^+$. By introducing the new matrices
$A^{n_{i+1}}_{a_i,a_{i+1}}=U_{n_L,a_{i+1}}$ and $B^{n_{i+1}}_{a_{i+1},a_{i+2}}
=(V^+)_{a_{i+1},n_R}$, one recovers a MPS in a mixed decomposition but with
an orthogonality center shifted to the right. For a time step $\Delta t$,
the algorithm is the following:
start with an MPS in the right-canonical form (only $B$ matrices),
apply the first evolution operator $\exp{-ih_{1,2}\Delta t}$ and restore
the MPS structure by a SVD. The orthogonality center has shifted to the
right. Before applying the second evolution operator $\exp{-ih_{3,4}\Delta t}$,
the orthogonality center needs to be shifted once more using (\ref{ShiftCenter}).
Repeat until the orthogonality center reaches the right edge of the chain.
All evolution operators $\exp{-ih_{i,i+1}\Delta t}$ with $i$ odd have been
applied and the MPS is formed only of left-canonical matrices $A^{n_i}$, apart
from the last one $B^{n_N}$ if $N$ is even. Proceed now with the even indices
in reverse order, i.e. start with $\exp{-ih_{{N-2},N-1}\Delta t}$ (if $N$ is
even), so that the orthogonality center shifts to the left. When all operators
have been applied, the MPS is again formed only of right-canonical matrices
$B^{n_i}$ apart from the first one $A^{n_1}$. These successive applications of
the 2-site evolution operators are represented diagrammatically on Fig.~\ref{fig4}.
Note that the original algorithm is not based on the mixed decomposition
(\ref{MPSMixte}) but on the Vidal representation where the product involves
both matrices and singular values at each site~\cite{Vidal}. A simple
{\tt python} implementation of the TEBD algorithm in this representation
is provided in~\cite{Hauschild}.

Most of the computational time consumed by the TEBD algorithm is spent
calculating SVD's. Since the time complexity of the SVD of a $n\times m$ matrix
scales as ${\cal O}(m^2n+n^2m+n^3)$, the CPU time required to restore the MPS
structure after the application of a 2-site evolution operation is of order
$(d\chi)^3$. As a consequence, the total CPU time for a time evolution up to
the time $t=n_{\rm iter}\Delta t$ scales as $n_{\rm iter}N(d\chi)^3$.
Moreover, each SVD is followed by a truncation to the $\chi$
largest singular values, so a systematic error accumulates
during the time evolution. Furthermore, entanglement has been shown to
increase during the unitary evolution~\cite{Muth}. As a consequence, a good
approximation of $\ket{\psi(t)}$ requires larger and larger bond dimensions
$\chi$ as $t$ increases. An empirical criterion for the reliability
of the simulation is that the total truncation error (\ref{TruncationErr})
should remain below $10^{-7}$. Finally, for completeness, one should mention that
the truncations performed after the SVD's may also break symmetries and
violate conservation laws. It is also difficult to extend the TEBD algorithm
to Hamiltonians with long-range interactions. These problems are partially
solved by the TDVP algorithm based on the projected Schr\"odinger equation
	\be i{d\over dt}\ket\psi={\cal P}H_S\ket\psi\ee
where ${\cal P}$ is the projector onto the tangent space of the
MPS~\cite{Haegeman,Vanderstraeten,Paeckel,Bauernfeind}.

\section{Time evolution of a Matrix Product Density Operator}
\subsection{Vectorization of Lindblad equation}\label{SecAlgo1}
Like Schr\"odinger equation (\ref{Schrodinger}), Lindblad equation
(\ref{Lindblad}) is a linear first-order differential equation. However,
it is an equation for the density matrix $\rho_S\in\big[{\cal H}^*
\big]^{\otimes N}\otimes{\cal H}^{\otimes N}$ rather than for a
quantum state $\ket\psi\in{\cal H}^{\otimes N}$. Nevertheless,
Lindblad equation can be put under the form of a Schr\"odinger equation
by using the Choi-Jamielowski isomorphism~\cite{Szankowski}. The latter
is a one-to-one map of $\big[{\cal H}^*\big]^{\otimes N}\otimes
{\cal H}^{\otimes N}$ onto the vector space ${\cal H}^{\otimes N}
\otimes {\cal H}^{\otimes N}$, sometimes called the Liouville space.
Both spaces have a dimension $d^{2N}$ for our chain of $N$ sites. In
a basis $\{\ket i\}$ of ${\cal H}^{\otimes N}$, any density matrix
$\rho_S$ is mapped onto a vector $\kket\rho$ defined as
	\be\rho_S=\sum_{i,j}\rho_{ij}\ket i\bra j
	\longrightarrow \kket\rho=\sum_{i,j}\rho_{i,j}
	\ket i\otimes\ket j.			 \label{Choi}\ee
Under this map, Lindblad equation (\ref{Lindblad}) becomes
	\be {d\over dt}\kket\rho={\cal L}\kket\rho	\label{LindbladChoi}\ee
with the Liouvillian operator
	\be\fl{\cal L}=-i(H_S\otimes\identity-\identity\otimes H_S)
	+\sum_k \gamma_k\Big[L_k\otimes L_k^+
	-{1\over 2}L_k^+L_k\otimes\identity
	-{1\over 2}\identity\otimes L_k^+L_k\Big].	\label{Liouvillian}\ee
The formal solution of this equation is
	\be \kket{\rho(t)}=e^{{\cal L}t}\kket{\rho(0)}.
	\label{SolLindbladChoi}\ee

\begin{figure}
    \centering
    \includegraphics[width=0.75\textwidth]{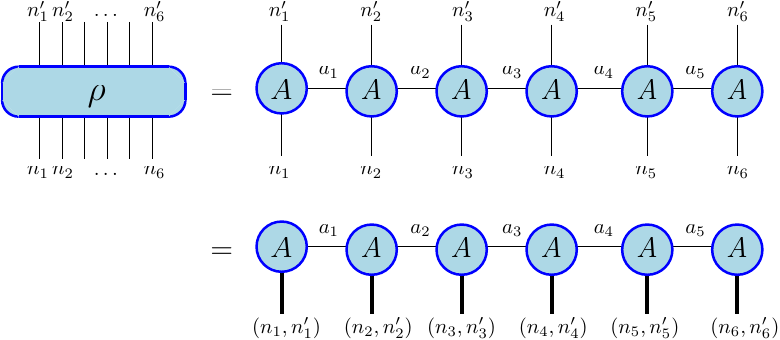}
    \caption{Diagrammatic representation of the density matrix (left)
    for a chain of 6 sites. Grouping together the indices $n_i$ and $n_i'$,
    successive SVD's yields the MPS depicted on the right (below).
    Alternatively, the density matrix can be put under the form of
    a product of matrices depending on the two indices $n_i$ and $n_i'$
    at site $i$ (above). The latter is referred to as a Matrix Product
    Operator (MPO) or a Matrix Product Density Operator (MPDO) to emphasize
    the fact that it corresponds to a density matrix.
    }\label{fig5}
\end{figure}

In the occupation-number basis, the decomposition of the quantum state
of the free fermion chain is given by (\ref{OccupationBasis}). In this
basis, Choi-Jamielowski isomorphism (\ref{Choi}) reads
	\ba\fl\rho_S=\sum_{n_1,\ldots,n_N,\atop n'_1,\ldots,n'_N}
	\rho_{n_1,\ldots,n_N;n'_1,\ldots,n'_N}
	\ket{n_1,\ldots,n_N}\bra{n'_1,\ldots,n'_N}	\nonumber\\
	\longrightarrow \kket\rho=\sum_{n_1,\ldots,n_N,\atop n'_1,
	\ldots,n'_N}\rho_{n_1,\ldots,n_N;n'_1,\ldots,n'_N}
	\ket{n_1,\ldots,n_N}\otimes\ket{n'_1,\ldots,n'_N}	\ea
A further reorganization of the indices leads to
	\be\kket\rho=\sum_{n_1,n'_1,\ldots\atop n_N,n'_N}
	\rho_{n_1,\ldots,n_N;n'_1,\ldots,n'_N}
	\ket{n_1}\otimes\ket{n'_1}\ldots
	\ket{n_N}\otimes\ket{n'_N}	\ee
which is similar to (\ref{OccupationBasis}) but with a local basis
$\{\ket{n_i}\otimes\ket{n'_i}\}$ on each site $i$ of the chain.
Grouping the two indices $n_i$ and $n'_i$ into a single one $m_i=n_id+n'_i$,
the vector $\kket\rho$ can be approached as a MPS (\ref{MPS})
	\be\rho_{n_1,\ldots,n_N;n'_1,\ldots,n'_N}=\sum_{a_1,\ldots a_{N-1}}
	A^{m_1}_{a_1}A^{m_2}_{a_1,a_2}\ldots A^{m_N}_{a_{N-1}}.\label{MPDO}\ee
The mapping is depicted on Fig.~\ref{fig5}. A TEBD algorithm based on the
Liouvillian (\ref{Liouvillian}) allows to estimate the time evolution of
the MPS associated to $\kket\rho$. Extensions of the {\tt iTensor} library
following this approach are available~\cite{Casagrande,Houdayer}. Note
that the same formalism allows the DMRG algorithm to be applied
to determine the steady-state~\cite{Cui}.
\\

Compared to the TEBD algorithm applied to a pure state, the only difference
is that the physical indices are now $m_i=n_id+n'_i$ and take therefore
$d^2$ values. It follows that the total CPU time after $n_{\rm iter}$ iterations
scales as $n_{\rm iter}N d^6\chi^3$. It may seem prohibitive if $d$ is large,
for instance for spin-$S$ quantum chains with $S$ large, but it is often argued
that dissipation destroys entanglement so the bond dimension $\chi$ may take
smaller values than for a unitary dynamics.

\def\DeltaT{0.003}

\begin{figure}
    \centering
    \includegraphics[width=0.5513\textwidth]{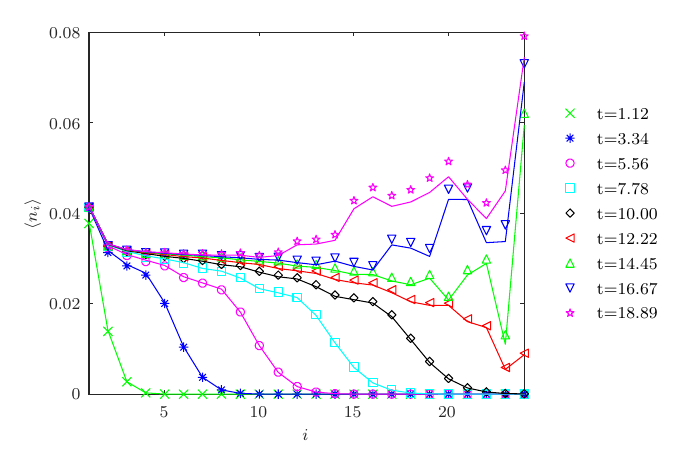}
    \includegraphics[width=0.4287\textwidth]{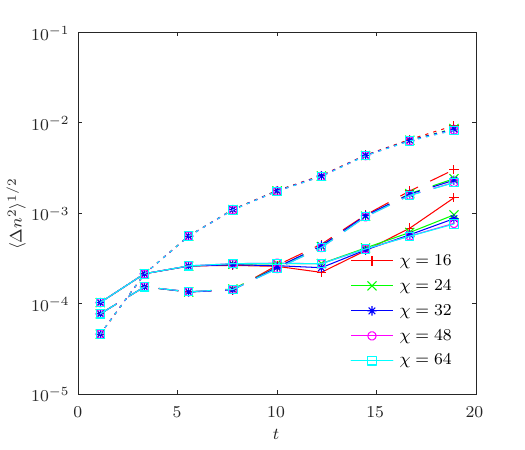}
    \caption{On the left, average density profile $\langle n_i\rangle$ of a
    free-fermion model with a particle source on the first site of the chain
    and a sink on the last one. The continuous lines correspond to the exact profile
    while the symbols were obtained numerically by vectorization of Lindblad
    equation and a MPDO representation with $\chi=64$ and $\Delta t=\DeltaT$.
    On the right, mean square deviation $\sqrt{{1\over N}
    \sum_i[\langle n_i\rangle-n_i^{\rm exact}]^2}$ of the density profile.
    The different symbols correspond to different maximal bond dimensions $\chi$
    and the different line styles to time steps $\Delta t=0.01$ (dotted),
    $0.003$ (dashed) and $0.001$ (continuous).
    }\label{fig-Algo1a}
\end{figure}

\begin{figure}
    \centering
    \includegraphics[width=0.5513\textwidth]{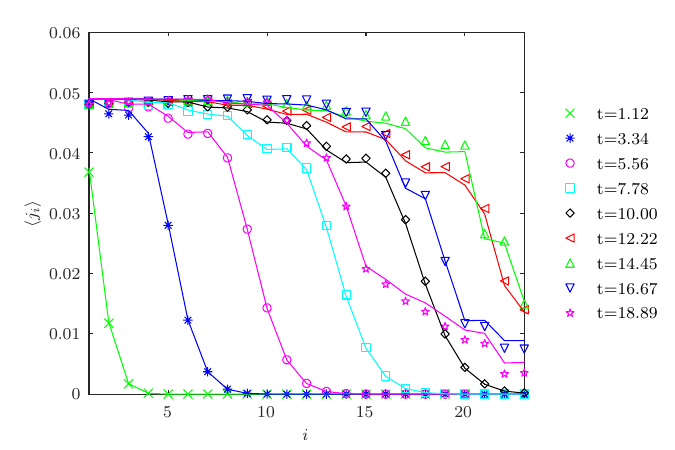}
    \includegraphics[width=0.4287\textwidth]{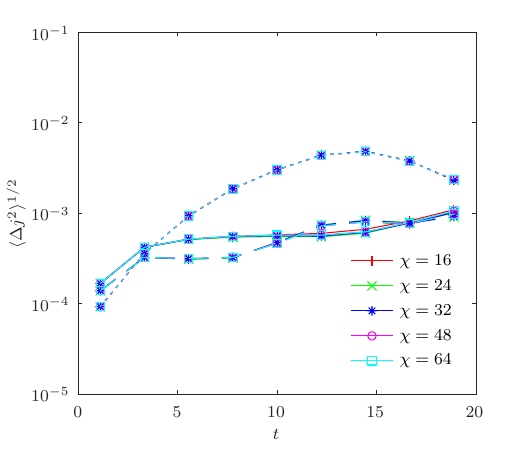}
    \caption{On the left, average current profile $\langle j_i\rangle$ of a
    free-fermion model with a particle source on the first site of the chain
    and a sink on the last one. The continuous lines correspond to the exact profile
    while the symbols were obtained numerically by vectorization of Lindblad
    equation and a MPDO representation with $\chi=64$ and $\Delta t=\DeltaT$.
    On the right, mean square deviation $\sqrt{{1\over N}
    \sum_i[\langle j_i\rangle-j_i^{\rm exact}]^2}$ of the current profile.
    The different symbols correspond to different maximal bond dimensions $\chi$
    and the different line styles to time steps $\Delta t=0.01$ (dotted),
    $0.003$ (dashed) and $0.001$ (continuous).
    }\label{fig-Algo1b}
\end{figure}

Using the above-described algorithm, simulations have been performed for the
free-fermion model with a particle source on the first site of the chain and
a sink on the last one presented at section~\ref{SecFermion}.
The comparison of the numerical estimates with
the exact values is shown on Fig.~\ref{fig-Algo1a} for the density profile
and on Fig.~\ref{fig-Algo1b} for the current. As can be seen on the plots on
the left, the agreement is relatively good for times up to $t=18.89$. On the
right, the average quadratic deviation of the numerical estimates, defined as
        \be\sqrt{{1\over N}\sum_i[\langle n_i(t)\rangle-n_i^{\rm exact}(t)]^2}\ee
for the density profile, is plotted for different maximal bond dimensions $\chi$ and
time steps $\Delta t$. At large times $t$, the deviation of the density
profile is mainly affected by the time step $\Delta t$. The influence of
the bond dimension $\chi$ is manifest only at the largest time $t=18.89$.
For the current, the deviation is more difficult to interpret. 

\subsection{Locally purification density matrix}\label{SecAlgo2}
\begin{figure}
    \centering
    \includegraphics[width=0.75\textwidth]{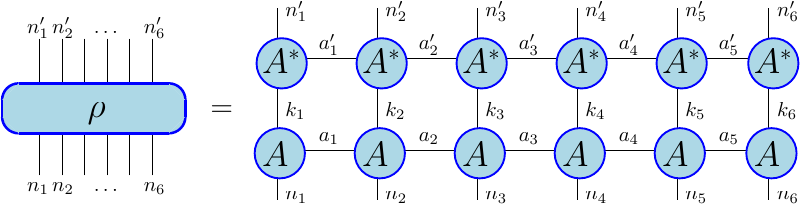}
    \caption{Diagrammatic representation of the locally purified
    density matrix (\ref{Purified2}) for a chain of 6 sites.
    }\label{fig6}
\end{figure}

Since the eigenvalues of the density matrix can be interpreted as probabilities,
they must be positive or zero, i.e. the density matrix is positive semi-definite.
This property is not enforced in the algorithm presented above so it may be
violated during time evolution. One possibility to impose positivity is to
write the density matrix as $\rho_S=XX^+$~\cite{Verstraete04,Zwolak}, which
corresponds to a mixed state. Indeed, in a basis $\{\ket i\}$ of
${\cal H}^{\otimes N}$, the density matrix reads explicitly
	\be\fl\rho_S=XX^+=\sum_{i,j,k} X_{ik}X^*_{jk}\ket i\bra j
	=\sum_k \Big[\sum_i X_{ik}\ket i\Big]\Big[\sum_j X_{jk}\ket j\Big]^+
	=\sum_k \wp_k\ket{\psi_k}\bra{\psi_k}	\label{Purified}\ee
where $\sqrt{\wp_k}\ket{\psi_k}=\sum_i X_{ik}\ket i$ and $\sqrt{\wp_k}$
is such that $\ket{\psi_k}$ is normalized. The size of $X$ being large,
$X$ will be approached by a MPO
	\be X_{n_1,\ldots n_N;k_1,\ldots k_N}=\sum_{a_1,\ldots a_{N-1}}
	A^{n_1,k_1}_{a_1}A^{n_2,k_2}_{a_1,a_2}\ldots A^{n_N,k_N}_{a_{N-1}}
	\label{Purified1}\ee
so that
	\ba\fl\rho_{n_1,\ldots n_N;n_1',\ldots n_N'}
	=\sum_{k_1,\ldots,k_N} X_{n_1,\ldots n_N;k_1,\ldots k_N}
	X^*_{n_1',\ldots n_N';k_1,\ldots n_N}				\nonumber\\
	\fl\quad =\sum_{a_1,\ldots a_{N-1},\atop a_1',\ldots a_{N-1}'}
	\bigg[\sum_{k_1} A^{n_1,k_1}_{a_1}(A^*)^{n_1',k_1}_{a_1'}\bigg]\ \!
	\bigg[\sum_{k_2} A^{n_2,k_2}_{a_1,a_2}(A^*)^{n_2',k_2}_{a_1,a_2'}\bigg]
	\ldots \bigg[\sum_{k_N} A^{n_N,k_N}_{a_{N-1}}
	(A^*)^{n_N',k_N}_{a_{N-1}'}\bigg]				\label{Purified2}\ea
A diagrammatic representation of this equality is shown on Fig.~\ref{fig6}.
$X$ does not need to be a square matrix so the internal indices $k_i$ can take
a different number of values than the $a_i$ and $a_i'$. While the dimension
of the horizontal bonds is $\chi$, that of the vertical internal ones
will be denoted $\chi_k$.

It now remains to extend the TEBD algorithm to generate the time evolution
of the locally purified density matrix (\ref{Purified}). In the following,
a simpler version of the algorithm presented in Ref.~\cite{Werner} is
discussed. As before, we will restrict ourselves to an Hamiltonian
$H_S=\sum_i h_{i,i+1}$ composed of terms acting only on two adjacent
sites. Moreover, it will be assumed that jump operators act only on one
site of the chain. The jump operator acting on site $i$ will be denoted $L_i$.
Lindblad equation (\ref{Lindblad}) can then be rewritten as
	\be {d\rho_S\over dt}=-i\sum_i h_{i,i+1}^{\rm eff}\rho_S
	+i\sum_i\rho_S\big(h_{i,i+1}^{\rm eff}\big)^+
	+\sum_i \gamma_i L_i\rho_S L_i^+			\label{LindbladLocal}\ee
with the effective non-hermitian 2-site Hamiltonians in the bulk ($1<i<N-1$)
	\be h_{i,i+1}^{\rm eff}=h_{i,i+1}-{i\over 4}\gamma_i L_i^+L_i
	-{i\over 4}\gamma_{i+1}L_{i+1}^+L_{i+1}\ee
and at the two edges of the chain
	\be\fl h_{1,2}^{\rm eff}=h_{1,2}-{i\over 2}\gamma_1 L_1^+L_1
	-{i\over 4}\gamma_2L_2^+L_2,\quad  h_{N-1,N}^{\rm eff}=h_{N-1,N}
	-{i\over 4}\gamma_{N-1}L_{N-1}^+L_{N-1}-{i\over 2}\gamma_NL_N^+L_N.\ee
The first two terms of (\ref{LindbladLocal}) are similar to von Neumann
equation (\ref{vonNeumann}) and generate the following time-evolution
	\be \rho_S(t)=e^{t{\cal L}_1}\rho_S(0)
	=e^{-i\sum_i h_{i,i+1}^{\rm eff}\ \!t}\rho_S(0)
	e^{i\sum_i (h_{i,i+1}^{\rm eff})^+\ \!t}.	\label{SolLindbladEff1}\ee
The density matrix being written as $\rho_S=XX^+$, one obtains
	\be \rho_S(t)=\Big[e^{-i\sum_i h_{i,i+1}^{\rm eff}\ \!t}X(0)\Big]
	\Big[e^{-i\sum_i h_{i,i+1}^{\rm eff}\ \!t}X(0)\Big]^+=X(t)X^+(t).\ee
It is therefore sufficient to compute the time-evolution of the matrix $X$,
or equivalently of all the states $\ket{\psi_k}$ of the mixed state
(\ref{Purified}). The matrix $X$ being expressed as a MPO (\ref{Purified1}),
the calculation of its time-evolution can be performed with the TEBD
algorithm for the non-hermitian 2-site Hamiltonian $h_{i,i+1}^{\rm eff}$.
Note that the latter does not involve the internal indices $k_i$ and
$k_{i+1}$ so that the matrix $\Theta$ (\ref{Theta}) should be defined as
	\be\fl\Theta_{a_{i-1},n_{i}',k_i,a_{i+1},n_{i+1}',k_{i+1}}
	=\sum_{a_{i},n_{i},n_{i+1}}\Big[e^{-ih_{i,i+1}\Delta t}
	\Big]_{n_{i}',n_{i+1}';n_{i},n_{i+1}}A^{n_{i},k_i}_{a_{i-1},a_{i}}
	A^{n_{i+1},k_{i+1}}_{a_{i},a_{i+1}}.	\label{ThetaPurified}\ee
The size of this matrix is now $(d\chi\chi_k)\times(d\chi\chi_k)$ so
its SVD will require a CPU time growing as $(d\chi\chi_k)^3$.
Note moreover that the norm of the wavefunction should not be enforced
at each iteration since the effective Hamiltonian is not hermitian.
\\

The last term of (\ref{LindbladLocal}) was omitted in the above discussion.
It displays a more complex structure than the first two ones, with $L_i$
acting on the left of $\rho_S$ and $L_i^+$ on the right, and generates
a time-evolution which is not of the simple form (\ref{SolLindbladEff1}).
However, after a Suzuki-Trotter decomposition, the two dynamics are
disentangled at first order in $\Delta t$:
	\be e^{\Delta t{\cal L}}\simeq e^{\Delta t{\cal L}_1}
	e^{\Delta t{\cal L}_2}.\ee
The assumption that $L_i$ acts only on site $i$, so that $[L_i,L_j]=0$,
brings the further simplification
	\be e^{\Delta t{\cal L}_2}=\prod_i e^{\gamma_i\Delta t
	L_i\otimes L_i^+}\simeq \prod_i \big(\identity+\gamma_i\Delta t
	L_i\otimes L_i^+\big)\ee
at first order in $\Delta t$. The density matrix being decomposed
as a product (\ref{Purified2}), the operator $\identity+\gamma_i\Delta t
L_i\otimes L_i^+$ acts only on the matrices $\sum_{k_i} A^{n_i,k_i}
(A^{n_i',k_i})^*$ and therefore preserves the structure of $\rho_S$ as
a product. The action of $\identity+\gamma_i\Delta t L_i\otimes L_i^+$
amounts to replace $\sum_{k_i} A^{n_i,k_i}(A^{n_i',k_i})^*$ in the
product (\ref{Purified2}) by the tensor
	\ba R_{n_i,a_{i-1},a_i;n_i',a_{i-1}',a_i'}
	&=&\sum_{k_i}\bigg[A^{n_i,k_i}_{a_i,a_{i-1}}
	\big(A^{n_i',k_i}_{a_i',a_{i-1}'}\big)^*\bigg.			\\
	&&\bigg.\quad
	+\bigg(\sum_{m_i} (L_i)_{n_i,m_i}A^{m_i,k_i}_{a_i,a_{i-1}}\bigg)
	\bigg(\sum_{m_i'} (L_i)_{n_i',m_i'}A^{m_i',k_i}_{a_i',a_{i-1}'}
	\bigg)^*\bigg].				\nonumber\label{MatrixR}\ea
The last step of the algorithm is to decompose this tensor into a product
of the form $\sum_{k_i} A^{n_i,k_i}(A^{n_i',k_i})^*$. After reshaping the
tensor $R$ into a $d\chi^2\times d\chi^2$ hermitian matrix,
$R$ is diagonalized
	\be R=\sum_k \lambda_k\ket{\phi_k}\bra{\phi_k},\ee
which requires a CPU time of order $d^3\chi^6$. Finally, to restore
the structure (\ref{Purified2}), the matrix $A^{n_i,k_i}$ is replaced by
the components of the eigenvector
	\be A^{n_i,k_i}_{a_{i-1},a_i}=\sqrt{\lambda_{k_i}}
	\braket{n_i,a_{i-1},a_i}{\phi_{k_i}}.	\label{NewPurified}\ee
Again, to avoid the exponential growth of the matrices, the latter are
truncated by keeping only the $\chi_k$ largest eigenvalues $\lambda_{k_i}$.
\\

In conclusion, the modifications to bring to the TEBD algorithm are rather
limited. First, one should add to each matrix of the MPS a new internal
index $k_i$. The 2-site local Hamiltonian $h_{i,i+1}$ is replaced by
an effective non-hermitian one which does not change this internal index.
Apart from these modifications, the MPS is evolved as in the unitary case.
After a full sweep of the chain, each matrix $A^{n_i,k_i}$ is successively
considered and replaced by (\ref{NewPurified}) after construction
and diagonalization of the associated matrix $R$ (\ref{MatrixR}).
The total CPU time of this algorithm behaves as $n_{\rm iter}N
\big({\cal O}(d^3\chi^3\chi_k^3)+{\cal O}(d^3\chi^6)\big)$
to be compared with $n_{\rm iter}Nd^6\chi^3$ for the vectorized
Lindblad equation without purification. Since $\chi$ and $\chi_k$ are
usually much larger than $d$, the constraint of a positive density matrix
therefore comes at a significant computational cost.

\begin{figure}
    \centering
    \includegraphics[width=0.5513\textwidth]{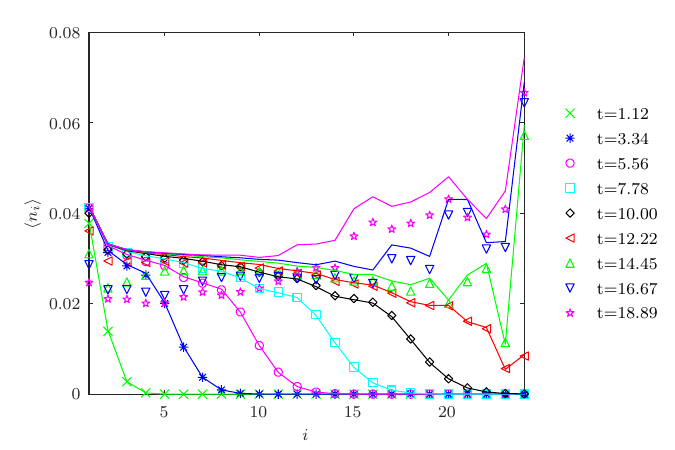}
    \includegraphics[width=0.4287\textwidth]{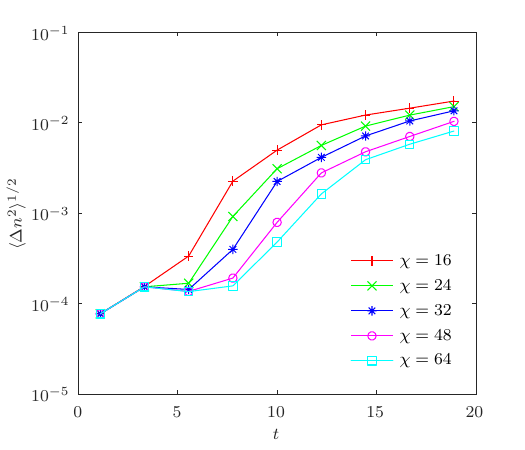}
    \caption{On the left, average density profile $\langle n_i\rangle$ of a
    free-fermion model with a particle source on the first site of the chain
    and a sink on the last one. The continuous lines correspond to the exact profile
    while the symbols were obtained numerically by vectorization of Lindblad
    equation and a purified MPDO representation with $\chi=\chi_k=64$
    and $\Delta t=\DeltaT$. On the right, mean square deviation $\sqrt{{1\over N}
    \sum_i[\langle n_i\rangle-n_i^{\rm exact}]^2}$ of the density profile.
    The different curves correspond to different maximal bond dimensions
    $\chi=\chi_k$.
    }\label{fig-Algo2a}
\end{figure}

\begin{figure}
    \centering
    \includegraphics[width=0.5513\textwidth]{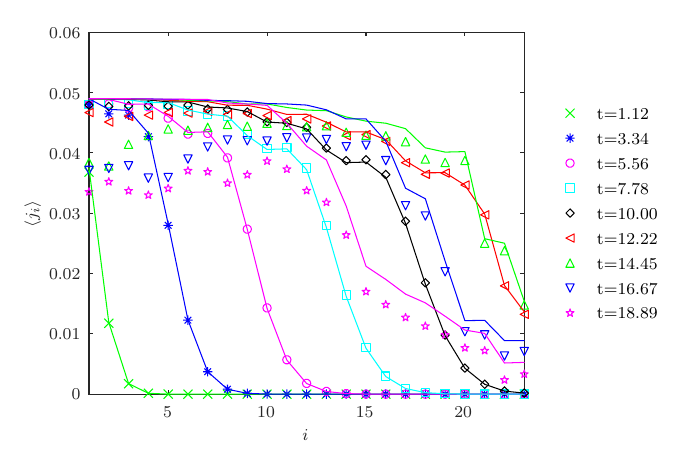}
    \includegraphics[width=0.4287\textwidth]{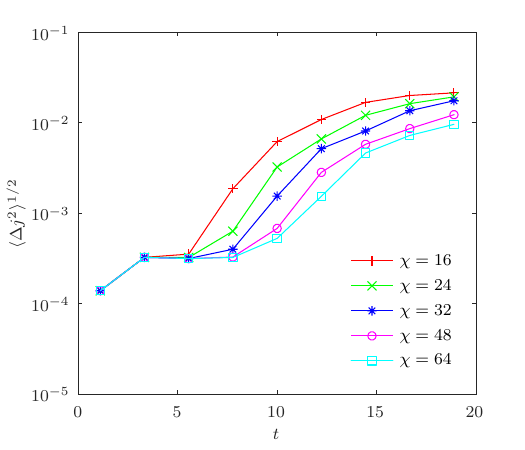}
    \caption{On the left, average current profile $\langle j_i\rangle$ of a
    free-fermion model with a particle source on the first site of the chain and
    a sink on the last one. The continuous lines correspond to the exact profile
    while the symbols were obtained numerically by vectorization of Lindblad
    equation and a purified MPDO representation with $\chi=\chi_k=64$
    and $\Delta t=\DeltaT$. On the right, mean square deviation $\sqrt{{1\over N}
    \sum_i[\langle j_i\rangle-j_i^{\rm exact}]^2}$ of the current profile.
    The different curves correspond to different maximal bond dimensions $\chi=\chi_k$.
    }\label{fig-Algo2b}
\end{figure}

Numerical results of simulations with this algorithm for the free-fermion model
with a particle source on the first site of the chain and a sink on the last
one are presented on figures \ref{fig-Algo2a} and \ref{fig-Algo2b}. For simplicity,
the two bond dimensions $\chi$ and $\chi_k$ were chosen equal. Although the
computation time is larger than for the vectorization algorithm of previous section,
the agreement of the numerical data with the data is much worse. At large times $t$,
a large deviation is observed. Moreover, on the first site of the chain, the density
and the current increase as expected up to $t\simeq 10$ but then, they both
decrease unexpectedly at larger times, even with a bond dimension $\chi=64$.
The average quadratic deviation plotted on the right of figures \ref{fig-Algo2a}
and \ref{fig-Algo2b} shows a strong dependence on the bond dimension $\chi$ at
intermediate times. In conclusion, this algorithm should be reserved to regimes
where the positivity of the density matrix is violated by the previous algorithm.

\section{Stochastic methods}
In this section, two algorithms based on a stochastic approach are presented.
The state of the system is described by a quantum state $\ket{\psi(t)}$
whose time evolution is not deterministic but stochastic. The latter is
constructed in a such a way that the average density matrix
	\be\overline{\rho_S(t)}=\overline{\ket{\psi(t)}\bra{\psi(t)}}\ee
evolves under Lindblad equation (\ref{Lindblad}). Numerically,
$\overline{\rho_S(t)}$ is approximated via a Monte Carlo simulation, i.e.
by a simple sampling of a large number $n_{\rm conf.}$ of histories
$\ket{\psi_\alpha(t)}$ of the system and an average of their density matrix
$\ket{\psi_\alpha(t)}\bra{\psi_\alpha(t)}$.
The first approach that will be discussed in this section, called
{\sl Quantum trajectories}, was originally introduced in the field of
quantum optics~\cite{Dum,Dalibard,Plenio,Daley}. Implementations with
a MPS representation of the state $\ket{\psi_\alpha(t)}$ are more
recent~\cite{Vovk1,Vovk2}. In this approach, the system undergoes
random quantum jumps to mimic the action of the last term of
(\ref{LindbladEff}). In the second approach that will be discussed,
called {\sl Quantum State Diffusion}, this action takes the form of
a classical noise coupled to the system~\cite{Chenu}.

\subsection{Quantum trajectories}\label{SecAlgo3}	
Consider the decomposition of the reduced density matrix
        \be\rho_S(t)=\sum_\alpha \wp_\alpha(t)\ket{\psi_\alpha(t)}
        \bra{\psi_\alpha(t)}   \label{AnsatzTrajQuantique}\ee
where the sum extends over all normalized vectors of the entire Hilbert space
${\cal H}^{\otimes N}$. The normalization condition $\trace\rho_S=1$ implies
that $\sum_\alpha\wp_\alpha(t)=1$, as expected for a probability. Note that
the decomposition (\ref{AnsatzTrajQuantique}) of the density matrix is not
unique. The expansion of $\rho_S$ over its instantaneous eigenbasis at time $t$,
with $\wp_\alpha(t)=0$ for all vectors that are not eigenvectors, is only one
of the possible decompositions.
\\

Both the probabilities $\wp_\alpha$ and the quantum states $\ket{\psi_\alpha}$
are assumed to be time-dependent. The first task is to determine
the equation of motion of $\wp_\alpha(t)$ and $\ket{\psi_\alpha(t)}$ so that
Lindblad equation (\ref{Lindblad}) is recovered. Instead of simply presenting
these equations and leaving to the reader to verify that Lindblad equation
holds, a heuristic and more intuitive derivation is proposed in the following.
Consider Lindblad equation under the form (\ref{LindbladEff}). It is tempting
to assume that the first two terms of the r.h.s. generate the time evolution
of the quantum states $\ket{\psi_\alpha(t)}$ while the last one comes from the
time-evolution of the probabilities $\wp_\alpha$. Our first guess is therefore
that the vectors $\ket{\psi_\alpha}$ evolve according to a Schr\"odinger
equation with the non-hermitian Hamiltonian (\ref{EffHamiltonian}), i.e.
        \be {d\over dt}\ket{\psi_\alpha}
        =-iH_{\rm eff}\ket{\psi_\alpha}
        =-i\Big(H-{i\over 2}\sum_k\gamma_k L_k^+L_k\Big)\ket{\psi_\alpha}.
        \label{QJump1}\ee
It follows that
		\ba\fl{d\over dt}\Big[\ket{\psi_\alpha}\bra{\psi_\alpha}\Big]
        &=&-i\Big(H-{i\over 2}
        \sum_k\gamma_k L_k^+L_k\Big)\ket{\psi_\alpha}\bra{\psi_\alpha}
        +i\ket{\psi_\alpha}\bra{\psi_\alpha}\Big(H+{i\over 2}
        \sum_k\gamma_k L_k^+L_k\Big)                            \nonumber\\
        \fl&=&-i[H,\ket{\psi_\alpha}\bra{\psi_\alpha}]
        -{1\over 2}\sum_k\gamma_k\{L_k^+L_k,\ket{\psi_\alpha}\bra{\psi_\alpha}\}.
        \ea
Multiplying by $\wp_\alpha$, summing over $\alpha$ and using
(\ref{AnsatzTrajQuantique}), one obtains
	\be {d\rho_s\over dt}=\sum_\alpha {d\wp_\alpha\over dt}
	\ket{\psi_\alpha}\bra{\psi_\alpha}-i[H,\rho_S]
	-{1\over 2}\sum_k\gamma_k\{L_k^+L_k,\rho_S\}\ee
from which one can hopefully deduce the equation of motion of the probabilities
$\wp_\alpha$ by identifying with Lindblad equation (\ref{Lindblad}). There is
however a flaw: because the effective Hamiltonian is not hermitian, the norm
of the states $\ket{\psi_\alpha}$ is not preserved by the equation of motion
(\ref{QJump1}). Indeed, one can check that
        \ba\fl{d\over dt}\braket{\psi_\alpha}{\psi_\alpha}
        &=&i\bra{\psi_\alpha}\Big(H+{i\over 2}\sum_k
        \gamma_k L_k^+L_k\Big)\ket{\psi_\alpha}
        -i\bra{\psi_\alpha}\Big(H-{i\over 2}
        \sum_k\gamma_k L_k^+L_k\Big)\ket{\psi_\alpha}    \nonumber\\
        \fl&=&-\sum_k\gamma_k \bra{\psi_\alpha}L_k^+L_k\ket{\psi_\alpha}\nonumber\\
        \fl&=&-\sum_k\gamma_k ||L_k\ket{\psi_\alpha}||^2.
        \label{DecroisPsiLindblad}\ea
To compensate this term and preserve the norm as assumed in (\ref{AnsatzTrajQuantique}),
the initial guess (\ref{QJump1}) is replaced by the equation of motion
	\be\fl{d\over dt}\ket{\psi_\alpha}\bra{\psi_\alpha}
        =-i[H,\ket{\psi_\alpha}\bra{\psi_\alpha}]
        -{1\over 2}\sum_k\gamma_k\{L_k^+L_k,\ket{\psi_\alpha}\bra{\psi_\alpha}\}
        +\sum_k\gamma_k ||L_k\ket{\psi_\alpha}||^2\ket{\psi_\alpha}\bra{\psi_\alpha}.
        \label{QJump2}\ee
By taking the trace of (\ref{QJump2}), one can check that the norm of
$\ket{\psi_\alpha}$ is now preserved. Multiplying now (\ref{QJump2}) by
$\wp_\alpha$, summing over $\alpha$ and using (\ref{AnsatzTrajQuantique}),
one obtains this time
	\be\fl{d\rho_S\over dt}=\sum_\alpha\bigg({d\wp_\alpha\over dt}
	+\sum_k\gamma_k||L_k\ket{\psi_\alpha}||^2\wp_\alpha\bigg)\ket{\psi_\alpha}
	\bra{\psi_\alpha}-i[H,\rho]-{1\over 2}\sum_k\gamma_k\{L_k^+L_k,\rho_S\}.\ee
and, by equating with Lindblad equation,
        \be\fl\sum_\alpha\bigg({d\wp_\alpha\over dt}+\sum_k\gamma_k
        ||L_k\ket{\psi_\alpha}||^2\wp_\alpha\bigg)
        \ket{\psi_\alpha}\bra{\psi_\alpha}
        =\sum_k\gamma_k L_k\rho_S L_k^+
        =\sum_{k,\beta} \gamma_k\wp_\beta L_k
        \ket{\psi_\beta}\bra{\psi_\beta}L_k^+.  \label{QJump3}\ee
Denote as $\ket{\psi_{\alpha_k}}$ the normalized vector such that
        \be{L_k\ket{\psi_{\alpha_k}}\over ||L_k\ket{\psi_{\alpha_k}}||}
        =\ket{\psi_\alpha}.\ee
Note that for a given state $\ket{\psi_\alpha}$ and a jump operator $L_k$,
$\ket{\psi_{\alpha_k}}$ may not exist. This is for the example the case in the
free fermion model for the vacuum $\ket{\psi_\alpha}=\ket 0$ when the
jump operators $L_k$ are creation operators. However, if $\ket{\psi_{\alpha_k}}$
exists, it is unique. Replacing the sum over $\beta$ in (\ref{QJump3}) by
a sum over $\alpha_k$, and then over $\alpha$, equation (\ref{QJump3}) becomes
        \be\sum_\alpha \left[{d\wp_\alpha\over dt}
        +\sum_k\gamma_k\wp_\alpha||L_k\ket{\psi_\alpha}||^2
        -\sum_k\gamma_k\wp_{\alpha_k}||L_k\ket{\psi_{\alpha_k}}||^2
        \right]\ket{\psi_\alpha}\bra{\psi_\alpha}.\ee
A sufficient, but not necessary, condition for this equation to hold is
		\be {d\wp_\alpha\over dt}
		=-\sum_k\wp_\alpha\gamma_k||L_k\ket{\psi_\alpha}||^2
        +\sum_k\wp_{\alpha_k}\gamma_k||L_k\ket{\psi_{\alpha_k}}||^2.
        \label{MasterEq}\ee
which can be interpreted as the master equation governing a stochastic
process in the Hilbert space~\cite{vanKampen,Wio,Garcia}. The system is
assumed to be in the state $\ket{\psi_\alpha(t)}$ with the probability
$\wp_\alpha(t)$ at time $t$ and the quantity
	\be W(\alpha_k\rightarrow\alpha)
	=\gamma_k||L_k\ket{\psi_{\alpha_k}}||^2\label{TransitionRate}\ee
is the rate of a transition from the state $\ket{\psi_{\alpha_k}}$ to
$\ket{\psi_\alpha}$. The second term of the r.h.s. of (\ref{MasterEq})
corresponds to the probability gained by $\ket{\psi_\alpha}$ per unit of
time due to the transitions from the states $\ket{\psi_{\alpha_k}}$
to $\ket{\psi_\alpha}$. The first term is interpreted as the loss of
probability of the state $\ket{\psi_\alpha}$ per unit of time due to
the transitions from $\ket{\psi_\alpha}$ to the states
$\ket{\psi_{\alpha^k}}$ defined as
        \be{L_k\ket{\psi_{\alpha}}\over ||L_k\ket{\psi_{\alpha}}||}
        =\ket{\psi_{\alpha^k}}\ee
if they exist. Note that, according to (\ref{DecroisPsiLindblad}), this
loss is precisely the decay rate of the square norm of the state under
the evolution governed only by the non-hermitian effective Hamiltonian.
The stochastic evolution of $\ket{\psi_\alpha(t)}$ is called a quantum
trajectory and the transitions quantum jumps.

\begin{figure}
    \centering
    \includegraphics[width=0.4\textwidth]{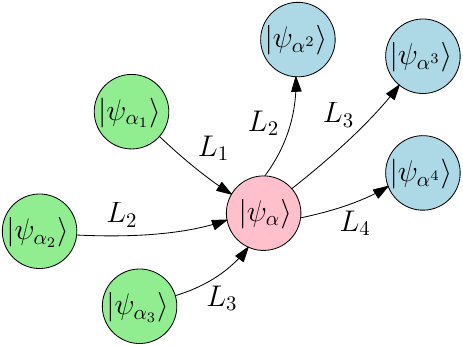}
    \caption{Representation of the stochastic process in Hilbert space
    governed by the master equation (\ref{MasterEq}). The possible transitions,
    represented by arrows, occur with the rate (\ref{TransitionRate}).
    The states from which a transition leads to $\ket{\psi_\alpha}$ are
    denoted by $\ket{\psi_{\alpha_k}}$ and the states to which a transition
    from $\ket{\psi_\alpha}$ can lead are denoted by $\ket{\psi_{\alpha^k}}$.
    In this example, there is no state $\ket{\psi_{\alpha_4}}$ such that
    $L_4\ket{\psi_{\alpha_4}}\sim\ket{\psi_{\alpha}}$ and no state
    $\ket{\psi_{\alpha^1}}$, i.e. $L_1\ket{\psi_{\alpha}}=0$.
    Note that for the free-fermion model, the two states
    $\ket{\psi_{\alpha^k}}$ and $\ket{\psi_{\alpha_{k'}}}$ may
    be equal if $L_k=c_i$ and $L_{k'}=c_i^+$ for example.
    }\label{fig-StochProcess}
\end{figure}

We now describe the numerical implementation. A single quantum state
$\ket{\psi_\alpha(t)}$ is considered and expressed as a MPS (\ref{MPS}).
Time is discretized and the time steps denoted $\Delta t$. At time $t=0$,
$\ket{\psi_\alpha}$ is chosen among the eigenvectors of the initial mixed
state $\rho_S(0)$ (\ref{AnsatzTrajQuantique}) with the probability distribution
$\wp_\alpha(0)$. At each time step $\Delta t$, all possible transition
rates $W(\alpha\rightarrow \alpha^k)$ are computed, which implies
to compute the norm of the states $L_k\ket{\psi_\alpha}$. If the MPS
is written in a mixed decomposition, this is hopefully a local operation
(see Fig.~\ref{fig3}). $\ket{\psi_\alpha}$ is replaced by one of the
states $\ket{\psi_{\alpha^k}}$ with the probability $W(\alpha\rightarrow
\alpha^k)\Delta t$. In the case where no transition occurred, which happens
with the probability $1-\sum_k W(\alpha\rightarrow\alpha^k)\Delta t$,
the state $\ket{\psi_\alpha}$ is evolved according to Schr\"odinger
equation (\ref{QJump1}). The TEBD algorithm with the effective Hamiltonian
$H_{\rm eff}$ can be employed. As seen above, the norm is not preserved.
To recover the dynamics generated by (\ref{QJump2}), it is sufficient to
normalize $\ket{\psi_{\alpha}(t+\Delta t)}$. The density matrix is recovered
only after averaging over all quantum trajectories. Averages are computed
as $\bigmoyenne{O(t)}=\bra{\psi_{\alpha}(t)}\hat O\ket{\psi_{\alpha}(t)}$
for a given quantum trajectory $\ket{\psi_{\alpha}(t)}$ and then averaged
over quantum trajectories:
        \be \overline{\bigmoyenne{O(t)}}
        =\trace \rho_S\hat O=\sum_\alpha \wp_\alpha\bra{\psi_{\alpha}(t)}
        \hat O\ket{\psi_{\alpha}(t)}.\ee
The number of terms in this sum being large, it is estimated by a
simple-sampling Monte Carlo simulation. The same simulation is repeated
$n_{\rm conf.}$ times and $\bra{\psi_{\alpha}(t)}\hat O\ket{\psi_{\alpha}(t)}$
is simply averaged over the $n_{\rm conf.}$ values. Since the histories
are independent, the central limit theorem implies that the fluctuations
on this estimate of $\overline{\bigmoyenne{O(t)}}$ decays as $1/\sqrt{n_{\rm conf.}}$.
The total computational times of this approach is $n_{\rm conf.}\times
n_{\rm iter}N(d\chi)^3$ to be compared with $n_{\rm iter}Nd^6\chi^3$ for
the vectorized Lindblad equation without purification. Note however that
$\chi$ may be smaller for the quantum trajectories approach because it
relies only on a MPS and not a MPDO so that the internal indices carry the
information about the correlations between degrees of freedom taking
$d$ values rather than $d^2$.
\\

A limitation of the quantum trajectory method combined to a state
$\ket{\psi_\alpha}$ described by a MPS is that the latter is numerically
optimized by the TEBD algorithm to describe accurately $\ket{\psi_\alpha}$
only. Therefore, $L_k\ket{\psi_\alpha}$ may be poorly described, especially
when $L_k$ acts as an annihilation operator. In the case of the free model
introduced in section~\ref{SecFermion}, we indeed observe a poor estimate
of the profile density close to the right edge of the chain where the
jump operator $L_2=c_N$ acts.
\\

To reduce the impact of quantum jumps, a common trick consists in shifting
the jump operators, i.e. to perform the replacement $L_k\longrightarrow L_k
+\Delta_k$ where $\Delta_k$ is a free parameter. Under this transformation,
the non-unitary part of Lindblad equation (\ref{Lindblad}) becomes
        \ba\fl\big(L_k+\Delta_k\big)\rho_S&&\big(L_k^++\Delta_k^*\big)
        -{1\over 2}\big(L_k^++\Delta_k^*\big)\big(L_k+\Delta_k\big)\rho_S
        -{1\over 2}\rho_S\big(L_k^++\Delta_k^*\big)\big(L_k+\Delta_k\big)\nonumber\\
        \fl&&=L_k\rho_S L_k^+-{1\over 2}\{L_k^+L_k,\rho_S\}
        +{1\over 2}\big[\big(\Delta_k^* L_k-\Delta_k L_k^+\big),\rho_S\big]
        \ea
To compensate for the last term and recover Lindblad equation,
$-{i\over 2}\sum_k\gamma_k\big(\Delta_k^*L_k-\Delta_k L_k^+\big)$
should be added to the Hamiltonian. The non-hermitian Hamiltonian becomes
therefore
        \be H_{\rm eff}=H-{i\over 2}\sum_k\gamma_k\big(L_k^+L_k+|\Delta_k|^2
        +2\Delta_k^*L_k\big).\ee
and the transition rates of the quantum jumps are
        \be W(\alpha_k\rightarrow\alpha)
        =\gamma_k||(L_k+\Delta_k)\ket{\psi_{\alpha_k}}||^2.\ee
To reduce the amplitude of quantum jumps, the optimal choice is $\Delta_k
=-\bra{\psi_\alpha(t)}L_k\ket{\psi_\alpha(t)}$. Under the action of the
jump operator $L_k+\Delta_k$, the wavefunction is then projected in a direction
perpendicular to $\ket{\psi_\alpha}$ since
        \be\bra{\psi_\alpha(t)}\big(L_k-\bigmoyenne{L_k}\big)
        \ket{\psi_\alpha(t)}=0\ee
with a transition rate proportional to the variance
        \be\bra{\psi_\alpha(t)}\big(L_k^+-\bigmoyenne{L_k}^*\big)
        \big(L_k-\bigmoyenne{L_k}\big)\ket{\psi_\alpha(t)}
        =\bigmoyenne{L_k^+L_k}-|\bigmoyenne{L_k}|^2.\ee

\begin{figure}
    \centering
    \includegraphics[width=0.5513\textwidth]{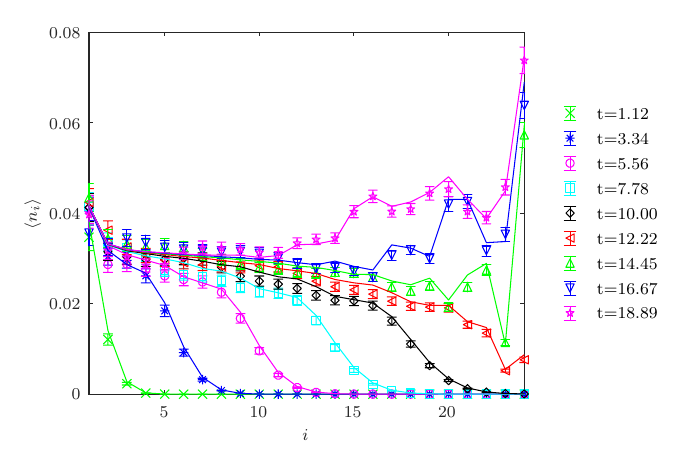}
    \includegraphics[width=0.4287\textwidth]{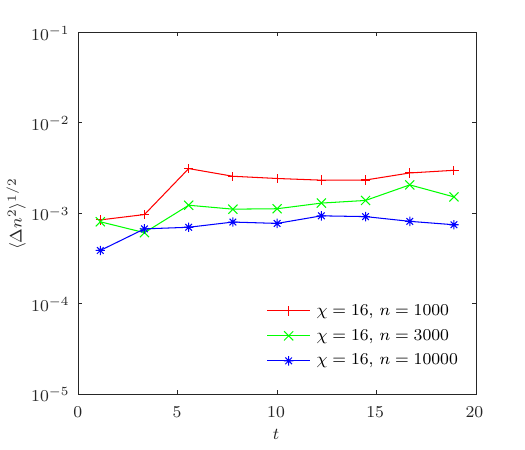}
    \caption{On the left, average density profile $\langle n_i\rangle$ of a
    free-fermion model with a particle source on the first site of the chain
    and a sink on the last one. The continuous lines correspond to the exact
    profile while the symbols were obtained numerically by a quantum trajectory
    algorithm based on a MPS representation with a maximal bond dimension $\chi=16$
    and $\Delta t=\DeltaT$. The data have been averaged over 3000 histories.
    On the right, mean square deviation $\sqrt{{1\over N}
    \sum_i[\langle n_i\rangle-n_i^{\rm exact}]^2}$ of the density profile.
    The different curves correspond to different numbers of trajectories $n_{\rm conf}$.
    }\label{fig-Algo3a}
\end{figure}

\begin{figure}
    \centering
    \includegraphics[width=0.5513\textwidth]{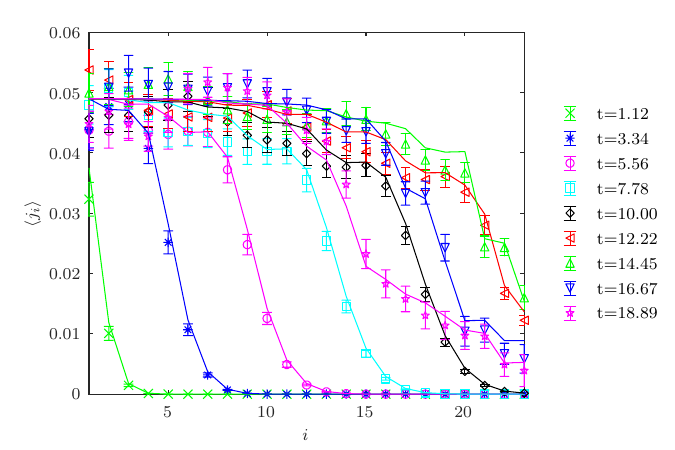}
    \includegraphics[width=0.4287\textwidth]{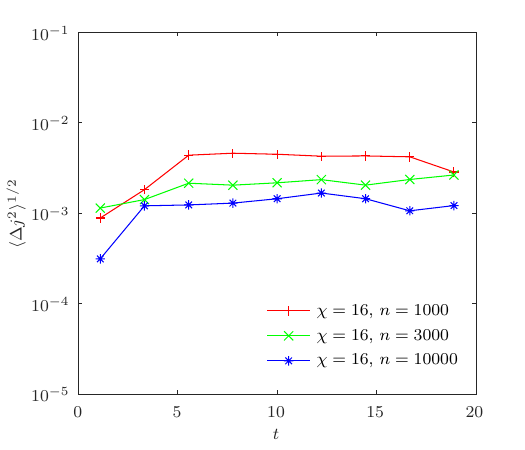}
    \caption{On the left, average current profile $\langle j_i\rangle$ of a
    free-fermion model with a particle source on the first site of the chain
    and a sink on the last one. The continuous lines correspond to the exact
    profile while the symbols were obtained numerically by a quantum trajectory
    algorithm based on a MPS representation with a maximal bond dimension $\chi=16$
    and $\Delta t=\DeltaT$. The data have been averaged over 3000 histories.
    On the right, mean square deviation $\sqrt{{1\over N}
    \sum_i[\langle j_i\rangle-j_i^{\rm exact}]^2}$ of the current profile.
    The different curves correspond to different numbers of trajectories $n_{\rm conf}$.
    }\label{fig-Algo3b}
\end{figure}

Numerical results of simulations with this algorithm for the free-fermion model
with a particle source on the first site of the chain and a sink on the last one
are presented on figures \ref{fig-Algo3a} and \ref{fig-Algo3b}. Error bars
on the average density and current correspond to the errors due to the finite
number of independent histories that were considered. According to the central
limit theorem, they can be estimated as
        \be \sqrt{\Delta n^2\over n_{\rm conf}}         \label{ErrorCLT}\ee
where $\Delta n^2$ is the variance of the density (or current).
Systematic deviations due to time discretization and to the truncation
of the matrices of the MPS are not taken into account into these errors.
Nevertheless, as can be observed on figures \ref{fig-Algo3a} and \ref{fig-Algo3b},
the data are compatible within errors bars with the exact density and current
profiles for most of the points. A better accuracy should be obtained with
a larger number of histories $n_{\rm conf.}$. The quadratic deviation indeed
shows a clear improvement when $n_{\rm conf.}$ is increased and almost none
with a larger bond dimension $\chi$.

\subsection{Quantum state diffusion algorithm}\label{SecAlgo4}
In the quantum trajectory approach, the dynamics of the wavefunction is
governed by Schr\"odinger equation for the effective non-hermitian Hamiltonian
and quantum jumps occurs randomly. For a system with low dissipation, the
latter are rare but the action of the jump operators on the wavefunction may be
poorly reproduced by a MPS. In this section, a different approach, based
on the equivalence of Lindblad equation with a stochastic process for
a pure state coupled to a classical noise~\cite{Gisin,Chenu}, is presented.
\\

Since the goal is to arrive at a numerical implementation, we start
immediately with a discrete time, the time step being denoted $\Delta t$.
This avoids recourse to the It\^o (or Stratonovich) formalism to deal with
stochastic equations. We assume that the time evolution of the wave
function is governed by an equation of the form
        \be\ket{\psi(t+\Delta t)}=\ket{\psi(t)}
        +\Big(\sum_k\eta_k A_k\sqrt{\Delta t}
        +B\Delta t\Big)\ket{\psi(t)}	\label{EqMvtStochPsi}\ee
where $\eta_k(t)$ are uncorrelated classical random variables satisfying
        \be\overline{\eta_k(t)}=0,\hskip 1truecm
        \overline{\eta_k(t)\eta_{k'}(t')}
        =\delta(t-t')\delta_{k,k'}.		\label{Noise}\ee
The operators $A_k$ and $B$ will be chosen in such a way that the
average density matrix $\overline{\rho_S(t)}=\overline{\ket\psi\bra\psi}$
satisfies Lindblad equation (\ref{Lindblad}). The time-evolution
of the density matrix follows from (\ref{EqMvtStochPsi}):
        \ba\fl&&\rho_S(t+\Delta t)
        =\Big[\identity+\Big(\sum_k\eta_k A_k\sqrt{\Delta t}
        +B\Delta t\Big)\Big]\ket{\psi(t)}\bra{\psi(t)}
        \Big[\identity+\Big(\sum_{k'}\eta_{k'} A_{k'}^+\sqrt{\Delta t}
		+B^+\Delta t\Big)\Big]
		\nonumber
        \label{DynStochLindblad1}\ea
where all random variables $\eta_k$ are taken at time $t$. Using the
moments (\ref{Noise}) of the noise, the average density matrix evolves as
        \be\overline{\rho_S(t+\Delta t)}=\overline{\rho_S(t)}
        +\Big[\sum_k A_k\overline{\rho_S(t)}A_k^+
        +B\overline{\rho_S(t)}+\overline{\rho_S(t)} B^+\Big]\Delta t
        +{\cal O}(\Delta t^2).\ee
Lindblad equation (\ref{Lindblad}) under the form
        \be\fl\overline{\rho_S(t+\Delta t)}=\overline{\rho_S(t)}
        +\bigg[-i[H,\overline{\rho_S(t)}]
        +\sum_k\gamma_k\bigg(L_k\overline{\rho_S(t)}L_k^+
        -{1\over 2}\{L_k^+L_k,\overline{\rho_S(t)}\}\bigg)\bigg]
        \Delta t+{\cal O}(\Delta t^2)\ee
is recovered by choosing
        \be A_k=\sqrt{\gamma_k}L_k,\hskip 1truecm
        B=-iH-\sum_k{\gamma_k\over 2}L_k^+L_k=-iH_{\rm eff}.	\label{DefAB}\ee
The equivalence with Lindblad equation ensures that the trace of the
average density matrix, and equivalently the average norm of the wavefunction,
is preserved. However, the trace of the density matrix and the norm
of the wavefunction fluctuate. Indeed, it can be checked that
		\ba\fl&&\braket{\psi(t+\Delta t)}{\psi(t+\Delta t)}\nonumber\\
        \fl&&=\bra{\psi(t)}\Big[\identity+\Big(\sum_k\eta_k A_k^+
        \sqrt{\Delta t}+B^+\Delta t\Big)\Big]
        \Big[\identity+\Big(\sum_{k'}\eta_{k'} A_{k'}\sqrt{\Delta t}
        +B\Delta t\Big)\Big]\ket{\psi(t)}        \nonumber\\
        \fl&&=\braket\psi\psi+\sum_k\eta_k\big(\bigmoyenne{A_k}
        +\bigmoyenne{A_k^+}\big)\sqrt{\Delta t}
        +\sum_{k,k'}\eta_k\eta_{k'} \bigmoyenne{A_k^+A_{k'}}\Delta t
		+\big(\bigmoyenne{B}+\bigmoyenne{B^+}\big)\Delta t
		+\ldots
        \label{DynStochLindblad2}\ea
where additional terms proportional to $\eta_k\Delta t^{3/2}$ have been
omitted. After averaging, the norm is conserved:
        \ba\fl
        \overline{\braket{\psi(t+\Delta t)}{\psi(t+\Delta t)}}
        &=&\overline{\braket\psi\psi}+\Big(\sum_k\bigmoyenne{A_k^+A_k}
        +\bigmoyenne{B}+\bigmoyenne{B^+}\Big)\Delta t+{\cal O}(\Delta t^2)
        \nonumber\\
        \fl&=&\overline{\braket\psi\psi}+\sum_k\gamma_k\big[\bigmoyenne{L_k^+L_k}
        -\bigmoyenne{L_k^+L_k}\big]\Delta t+{\cal O}(\Delta t^2)  \nonumber\\
        \fl&=&\overline{\braket\psi\psi}+{\cal O}(\Delta t^2)
        \ea
\\

The numerical implementation is rather straightforward. According to
(\ref{EqMvtStochPsi}) and (\ref{DefAB}), it amounts to iterate
the relation
		\ba\ket{\psi(t+\Delta t)}&=&\ket{\psi(t)}
        +\Big(\sum_k\eta_k\sqrt{\gamma_k\Delta t}L_k
        -iH_{\rm eff}\Delta t\Big)\ket{\psi(t)}	\nonumber\\
        &=&\prod_k\big(\identity+\eta_k\sqrt{\gamma_k\Delta t}L_k\big)
        \big(\identity-iH_{\rm eff}\Delta t)\ket{\psi(t)}
        +{\cal O}(\Delta t^{3/2})\ea
The wavefunction being written as a MPS, the application of $(\identity
-iH_{\rm eff}\Delta t)$ can be performed using the TEBD algorithm.
In the case where the jump operators $L_k$ are local and act only one site,
a full sweep of the chain is then performed and $\identity
+\eta_k\sqrt{\gamma_k\Delta t}L_k$ is applied to the matrix at the
orthogonality center. The MPS structure is preserved. Note that (\ref{Noise})
are the only constraints on the noise. Instead of a Gaussian noise, which
may seen the natural choice, we chose a binary distribution $\eta_k=\pm 1$
with a probability $1/2$. As for the quantum trajectories method,
quantum averages $\bra{\psi(t)}\hat O\ket{\psi(t)}$ needs to
be averaged over $n_{\rm conf.}$ histories of the system. Fluctuations
scale as $1/\sqrt{n_{\rm conf.}}$ while the computation time increases as
$n_{\rm conf.}\times n_{\rm iter}N(d\chi)^3$.
\\

The shift of the jump operators, previously discussed in the case of quantum
trajectories, can be implemented by choosing
	\be\fl A_k=\sqrt{\gamma_k}\big(L_k-\bigmoyenne{L_k}\big),\hskip 1truecm
	B=-iH-\sum_k\gamma_k\Big({1\over 2}L_k^+L_k-\bigmoyenne{L_k^+}L_k
	+{1\over 2}\bigmoyenne{L_k^+}\bigmoyenne{L_k}\Big).\ee
This choice leads to a reduction of the fluctuations of the norm
of the wavefunction. Since $\bigmoyenne{A_k}+\bigmoyenne{A_k^+}=0$, the term
proportional to $\sqrt{\Delta t}$ cancels in (\ref{DynStochLindblad2}).
A full calculation gives
	\be\fl\braket{\psi(t+\Delta t)}{\psi(t+\Delta t)}
	=\braket\psi\psi+\sum_{k\ne k'}\eta_k\eta_{k'}\sqrt{\gamma_k\gamma_{k'}}
	\bigmoyenne{\big(L_k^+-\bigmoyenne{L_k^+}\big)
	\big(L_{k'}-\bigmoyenne{L_{k'}}\big)}\Delta t\ee
at lowest order in $\Delta t$. A further improvement may be obtained
by replacing the first-order numerical scheme presented above for the
integration of Lindblad equation by higher-order schemes. However, the
second order scheme requires to double the number of random variables
which leads to larger fluctuations of the density matrix.

\begin{figure}
    \centering
    \includegraphics[width=0.49\textwidth]{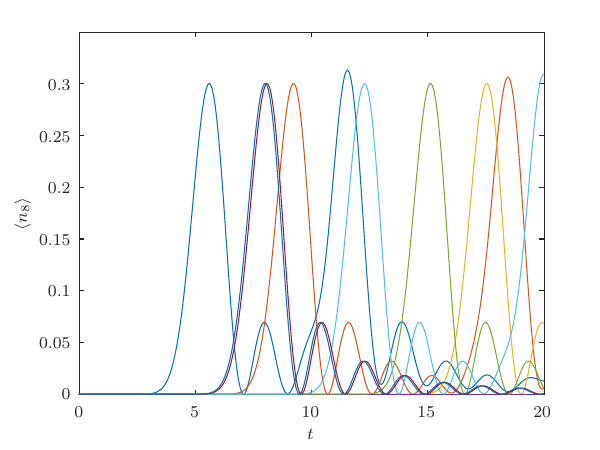}
    \includegraphics[width=0.49\textwidth]{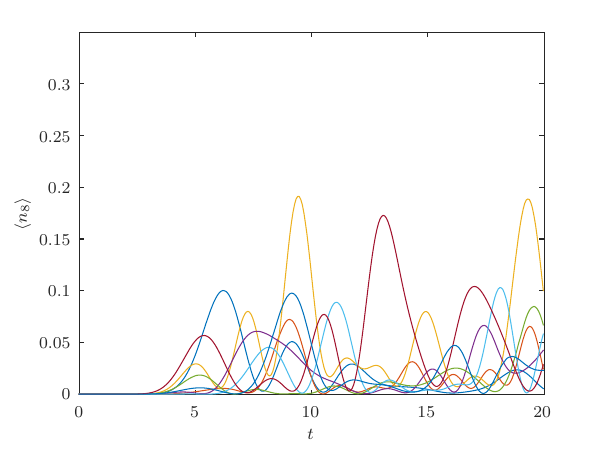}
    \caption{Time evolution of the average density $\langle n_8\rangle$
    on the 8-th site of the chain for eight randomly-chosen histories of
    the system computed with the quantum trajectory (left) and
    the quantum-state diffusion algorithms (right).}
    \label{fig-Algo5}
\end{figure}

\begin{figure}
    \centering
    \includegraphics[width=0.5513\textwidth]{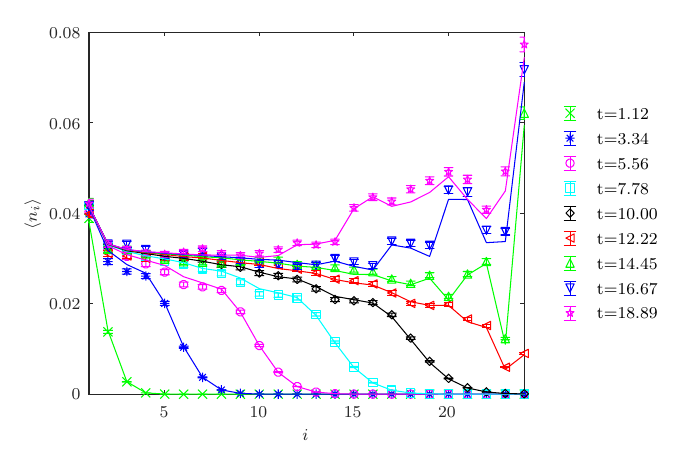}
    \includegraphics[width=0.4287\textwidth]{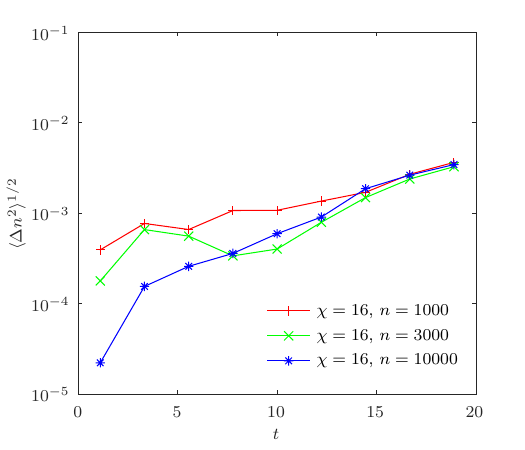}
    \caption{On the left, average density profile $\langle n_i\rangle$ of a
    free-fermion model with a particle source on the first site of the chain
    and a sink on the last one. The continuous lines correspond to the exact profile
    while the symbols were obtained numerically by a quantum state diffusion
    algorithm based on MPS representation with a bond dimension $\chi=16$
    and $\Delta t=\DeltaT$. The data have been averaged over 3000 histories.
    On the right, mean square deviation $\sqrt{{1\over N}
    \sum_i[\langle n_i\rangle-n_i^{\rm exact}]^2}$ of the density profile.
    The different curves correspond to different numbers of trajectories $n_{\rm conf}$.
    }\label{fig-Algo4a}
\end{figure}

\begin{figure}
    \centering
    \includegraphics[width=0.5513\textwidth]{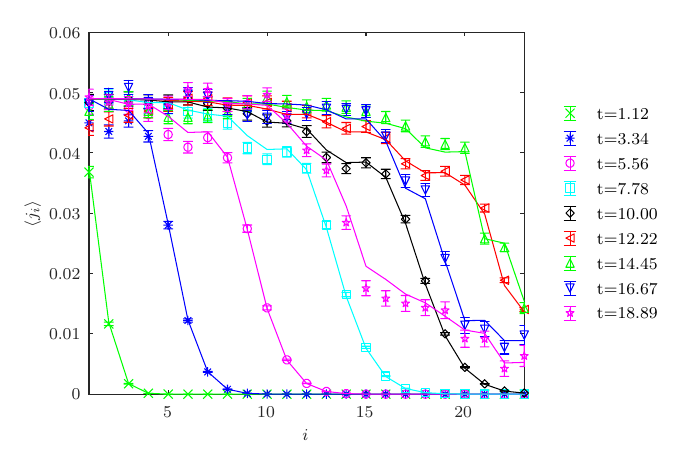}
    \includegraphics[width=0.4287\textwidth]{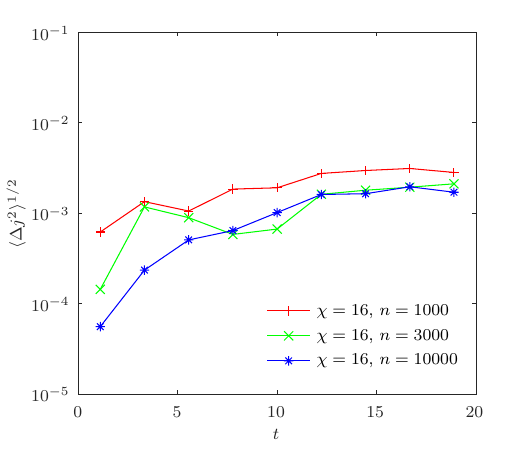}
    \caption{On the left, average current profile $\langle j_i\rangle$ of a
    free-fermion model with a particle source on the first site of the chain
    and a sink on the last one. The continuous lines correspond to the exact profile
    while the symbols were obtained numerically by a quantum state diffusion
    algorithm based on MPS representation with a bond dimension $\chi=16$
    and $\Delta t=\DeltaT$. The data have been averaged over 3000 histories.
	On the right, mean square deviation $\sqrt{{1\over N}
    \sum_i[\langle j_i\rangle-j_i^{\rm exact}]^2}$ of the current profile.
    The different curves correspond to different numbers of trajectories $n_{\rm conf}$.
    }\label{fig-Algo4b}
\end{figure}

Examples of independent histories obtained by the quantum trajectory and
quantum-state diffusion algorithms are shown on Fig.~\ref{fig-Algo5}. The
density on the 8-th site of the chain is plotted versus time. For the
quantum trajectory algorithm, one observes a sequence of large pulses
travelling through this site. Many of these pulses have an amplitude
as large as roughly $0.3$. In contrast, peaks are smaller but more
frequent for the quantum-state diffusion algorithm. As a consequence,
the average density displays smaller fluctuations for the quantum-state
diffusion algorithm, as can be observed on figures~\ref{fig-Algo4a}
and \ref{fig-Algo4b} for the current. Error bars, computed as
(\ref{ErrorCLT}), are indeed much smaller than on Fig.~\ref{fig-Algo3a}
and \ref{fig-Algo3b}. However, the exact density and current profile
are now outside error bars at large times. This indicates that
time discretization and truncation of the matrices of the MPS become
the largest source of error. Indeed, on the figures on the right, one
observes that the quadratic deviation tends to collapse for different
number of histories at large times, in contrast to the quantum trajectory
algorithm.

\section{Conclusions}
Several conclusions can be drawn from the simulations presented in this paper.
For comparable computational resources, the least accurate results have been
obtained with the algorithm based on a locally purified density matrix
(section~\ref{SecAlgo2}). As already noticed in the literature~\cite{Cuevas},
the computation cost of imposing the positivity of the density matrix
appears can be very high. The good results obtained with
the algorithm based on a MPDO (section~\ref{SecAlgo1})
suggest that, for the simple free-fermion model with a source and a sink,
the violation of the positivity of the density matrix remains small
in the time interval considered. It would have been interesting to check
this violation but it is known to be a computationally difficult task
in general~\cite{Kliesch}. For the model and parameters considered in this
study, the more sophisticated locally purified approach appears unnecessary.
This may, however, change for more complex models or at longer times, where
violation of positivity may become significant.
\\

The second conclusion that can be drawn is that, for the same bond dimension
$\chi$ and number of histories $n_{\rm conf.}$, the quantum-state diffusion
algorithm (section~\ref{SecAlgo4}) outperforms the quantum trajectory algorithm
(section~\ref{SecAlgo3}) at small times $t\lesssim 15$. However, while the
deviation to the exact results remains relatively stable with time $t$ for the
quantum trajectory algorithm, it increases rapidly for the quantum-state
diffusion algorithm. For a time $t\simeq 15$, the deviation is roughly the
same for the two algorithms. Figure~\ref{fig-Algo4b} shows that
time discretization and MPS truncation, rather than statistical sampling,
become the dominant sources of error at large times. Increasing the number
of histories alone does not significantly improve the accuracy in this regime.
\\

Finally, one should note that the computation time dedicated in this study
to the stochastic approaches (sections~\ref{SecAlgo3} and \ref{SecAlgo4})
was much larger than the one for the algorithm based on the vectorization
of the Lindblad equation (section~\ref{SecAlgo1}).
Within the present benchmark, the latter is therefore the most
efficient approach. This conclusion should nevertheless be regarded
as dependent on the model. It indeed holds for the 24-site free-fermion
model with a particle source and a sink that has been considered in this
study but it cannot be expected to be a general rule. Indeed, it has been
observed in the literature that the relative efficiency of different
numerical approaches can depend strongly on the form of the jump operators
considered~\cite{Kollath}. A more in-depth comparison should also involve
a study of the entanglement entropy growth to monitor the efficiency
of the MPS representation, as was performed in~\cite{Pressier} for the
MPDO (section~\ref{SecAlgo1}) and quantum trajectory algorithms
(sections~\ref{SecAlgo3}). It should also be noted that only the simplest
implementations of the various algorithms have been presented in this
introductory review. In particular, the order of the numerical scheme
is limited to the lowest order in $\Delta t$. More elaborated algorithms
may lead to drastic improvements in the accuracy.

\section*{Acknowledgments}
The author would like to thank Jean-Christophe Tremblay and Saad Yalouz
for the enriching and refreshing discussions.
This work was supported by the french ANR-PRME UNIOPEN grant (ANR-22-CE30-0004-01).

\section{Bibliography}

\def\book#1#2#3#4{#1 (#2) {\sl #3}, #4}
\def\paper#1#2#3#4#5{#1 (#2) {\sl #3} {\bf #4} #5}
\def\doi#1#2#3#4#5#6{#1 (#2) {\sl #3} {\bf #4} #5, doi:#6}
\def\preprint#1#2#3#4{#1 (#2) arXiv:#4}

\end{document}